\documentclass[12pt,draftcls,onecolumn]{IEEEtran}

\usepackage{amsfonts}
\usepackage{cite}
\usepackage{graphicx}
\usepackage{amssymb}
\usepackage{subfigure}
\usepackage{array}
\usepackage{color}
\usepackage{amsmath}
\usepackage{algorithmic}
\usepackage{algorithm}
\usepackage{cases}
\usepackage{bm}
\usepackage{mathrsfs}
\usepackage{epsfig}
\usepackage{psfrag}
\usepackage{url}
\usepackage{setspace}

\usepackage{etoolbox}
\makeatletter
\patchcmd{\@makecaption}
  {\scshape}
  {}
  {}
  {}
\makeatletter
\patchcmd{\@makecaption}
  {\\}
  {.\ }
  {}
  {}
\makeatother

\newcommand{\dis}{\stackrel{d}{\sim}}
\newcommand{\eqla}{\stackrel{(a)}{=}}
\newcommand{\eqlb}{\stackrel{(b)}{=}}

\newtheorem{Thm}{Theorem}
\newtheorem{Lem}{Lemma}
\newtheorem{Cor}{Corollary}

\newtheorem{Prob}{Problem}

\IEEEoverridecommandlockouts

\begin{document}

\title{Analysis and Optimization of Caching and Multicasting for Multi-Quality Videos in Large-Scale Wireless Networks}

\author{
\IEEEauthorblockN{Dongdong Jiang, {\em Student Member, IEEE} and Ying Cui, {\em Member, IEEE}} 
\thanks{D. Jiang and Y. Cui are with the Department of  Electronic Engineering, Shanghai Jiao Tong University, China.} 
}





\maketitle

\vspace{-8mm}
\begin{abstract}
Efficient dissemination of videos is an important problem for mobile telecom carriers.
In this paper, to facilitate massive video dissemination, we study joint caching and multicasting for multi-quality videos encoded using two video encoding techniques, namely scalable video coding (SVC), and HEVC or H.264 as in dynamic adaptive streaming over HTTP (DASH) respectively, in a large-scale wireless network. 
First, for each type of videos, we propose a random caching and multicasting scheme, carefully reflecting the relationship between layers of an SVC-based video or  descriptions of a DASH-based video. 
Then, 
for each type of videos, 
we derive tractable expressions for the successful transmission probability in the general and high user density regions, respectively, utilizing tools from stochastic geometry. The analytical results reveal that in the high user density region, the marginal increase of the successful transmission probability with respect to the caching probability of a video with a certain quality reduces when the caching probability increases. 
Next, for each type of videos, we consider the maximization of the successful transmission probability in the high user density region,  
which is a convex problem with an exceedingly large number of optimization variables.
We propose a two-stage optimization method to obtain a low-complexity near optimal solution by solving a relaxed convex problem and a related packing problem. The optimization results reveal the impact of the caching gain of a layer for an SVC-based video or a description for a DASH-based video on its caching probability. 
Finally, we show that  the proposed solutions for SVC-based and DASH-based videos achieve significant performance gains over baseline schemes in the general and high user density regions, and demonstrate their respective operating regions, using numerical results based on real video sequences.
\end{abstract}



\section{Introduction}

Mobile video traffic will occupy almost $75\%$ of the entire data traffic by 2020~\cite{cisco2016global}, posing a severe challenge for mobile telecom carriers.
Motivated by the fact that duplicate downloads of a small number of popular videos often account for a large portion of mobile data traffic~\cite{Yu06}, recently, caching popular videos at the wireless edge, namely caching helpers (base stations and access points) has been proposed as a promising approach for reducing delay and backhaul load.

Due to the nonuniform cellular usage costs, display resolutions of devices and channel conditions, users may have different quality requirements for a video.
Scalable video coding (SVC) can be used to create multi-quality videos. Specifically, SVC encodes each video into one base layer and multiple enhancement layers~\cite{Schwarz07}.
The base layer carries the essential information and provides a minimum quality of the video, and the enhancement layers represent the same video with gradually increasing quality. The decoding of a higher enhancement layer is based on the base layer and all its lower enhancement layers.
In~\cite{Poularakis16INFOCOM,Zhan17Lett,Ye17ITC,Zhang17WCNC,Yu16ICC,Tong16MSN}, the authors consider optimal caching designs for SVC-based videos to minimize the average delay~\cite{Poularakis16INFOCOM,Zhan17Lett}, the backhaul load~\cite{Ye17ITC} and the cache miss probability~\cite{Zhang17WCNC}, or to maximize the total quality levels of served video requests~\cite{Yu16ICC} and the quality of experience~\cite{Tong16MSN}. 
The optimization problems in~\cite{Poularakis16INFOCOM,Zhan17Lett,Ye17ITC,Zhang17WCNC,Yu16ICC,Tong16MSN} are NP-hard and heuristic approximate solutions are proposed.
Note that \cite{Poularakis16INFOCOM,Zhan17Lett,Ye17ITC,Zhang17WCNC,Yu16ICC,Tong16MSN} consider simple network models which cannot capture  geographic features of the locations of helpers and users or statistical properties of signals and interferences.
In~\cite{Zhang17GLOBECOM} and \cite{Wu17Lett}, more realistic large-scale network models based on stochastic geometry are considered to characterize stochastic natures of channel fading and geographic locations of helpers and users. Based on these models, the authors analyze the performance of some simple caching designs for SVC-based videos. Specifically, in \cite{Zhang17GLOBECOM}, 
the $\ell$-th nearest helper in the serving cluster stores the \textcolor{black}{base} layer and the $\ell$-th enhancement layer of each popular SVC-based video, and  the local service probability, ergodic service rate and service delay are analyzed. In \cite{Wu17Lett}, the authors consider SVC-based videos with two quality levels, store the base layer and the enhancement layer \textcolor{black}{of} each of the most popular SVC-based videos and only the base layer \textcolor{black}{of} each of the less popular SVC-based videos at each helper, 
and analyze the service delay. 
Note that 
the simple caching designs for SVC-based videos in \cite{Zhang17GLOBECOM} and \cite{Wu17Lett} may not make full use of network storage resource or achieve desirable performance. Optimal caching design for SVC-based videos in a large-scale wireless network is still not known.

To provide multi-quality video service, each video can also be encoded into multiple separate descriptions, each \textcolor{black}{of a certain} quality level, 
using HEVC or H.264 as in dynamic adaptive streaming over HTTP (DASH). 
The decoding of a description is independent of \textcolor{black}{that of any other description}~\cite{Park18TVT}.
In \cite{Araldo16IFIP}, the authors consider joint optimization of caching and routing for DASH-based videos to maximize the overall user utility in a simple network model.
In \cite{Choi18JSAC}, the authors analyze and optimize the performance of random caching for DASH-based videos in a large-scale wireless network. Note that \cite{Choi18JSAC} does not properly consider the interference in the network, and does not provide an achievable caching scheme corresponding to the optimal solution.

In \cite{Hartanto2006} and \cite{Poularakis14INFOCOM}, the authors compare caching designs for SVC-based videos and DASH-based videos in simple network models. Specifically, \cite{Hartanto2006} considers videos with two quality levels and adopts simple popularity-aware caching designs for the two types of videos;  
\cite{Poularakis14INFOCOM} considers the optimization of caching designs for the two types of videos 
and obtains heuristic solutions.
\textcolor{black}{Thus, the adopted caching designs in \cite{Hartanto2006} and \cite{Poularakis14INFOCOM}} may not fully exploit storage resource. It is still not known which video encoding technique can achieve better performance \textcolor{black}{in} cache-enabled wireless networks. 
\textcolor{black}{In addition}, enabling multicast service at helpers is an efficient way to deliver popular videos to multiple requesters simultaneously by effectively utilizing the broadcast nature of the wireless medium.
In~\cite{Chuah12TM} and \cite{Choi15TCOM}, the authors \textcolor{black}{study} optimal multi-quality multicast for SVC-based videos \textcolor{black}{using} simple fixed wireless network topologies.
In \cite{kim01comparison}, the authors compare the performance of multicasting SVC-based videos and \textcolor{black}{that of} DASH-based videos in wired networks. 
As far as we know, there \textcolor{black}{has been} no analysis or optimization results on multi-quality multicast for SVC-based or DASH-based videos in large-scale wireless networks.

It has been well recognized that jointly considering caching and multicasting can improve efficiency for content dissemination in wireless networks.
Our \textcolor{black}{previous} work~\cite{Cui16TWC} is the first one on the analysis and optimization of caching and multicasting for independent single-quality files in large-scale wireless networks. 
However, the results in~\cite{Cui16TWC} no longer hold for multi-quality files, such as SVC-based videos and DASH-based videos, as different layers of an SVC-based video and different descriptions of a DASH-based video have certain relations and cannot be treated as independent single-quality files. It is \textcolor{black}{of great importance} to understand how caching and multicasting can maximally improve the efficiency for disseminating SVC-based videos and DASH-based videos respectively in large-scale wireless networks, and which video encoding technique can achieve better performance when both caching and multicasting are considered. 

In this paper, we shall tackle the aforementioned issues. We consider joint caching and multicasting for SVC-based videos and DASH-based videos respectively in a large-scale wireless network. Our main contributions are summarized below.
\begin{itemize}
\item First, for each type of videos, we propose a random caching and multicasting scheme,  
carefully reflecting the relationship between layers of an SVC-based video or  descriptions of a DASH-based video. 
Each scheme can effectively adapt to the popularity of multi-quality videos, wisely exploit the storage resource, and fairly utilize the bandwidth resource. 

\item Then, for each type of videos, 
    we derive a tractable expression for the successful transmission probability, 
    utilizing tools from stochastic geometry. The expression reveals impacts of physical layer parameters and the caching distribution on the successful transmission probability.
    To obtain more design insights, we also obtain a closed-form expression for the successful transmission probability in the high user density region, by adopting some approximations. The expression reveals that in the high user density region, the marginal increase of the successful transmission probability with respect to the caching probability of 
    a video with \textcolor{black}{a} certain quality reduces when the caching probability increases. 

\item Next, for each type of videos, we consider the maximization of the successful transmission probability in the high user density region, which 
is a convex problem with an exceedingly large  number of optimization variables and  prohibitively high \textcolor{black}{computational} complexity.
We propose a two-stage optimization method to obtain a low-complexity near optimal solution. Specifically, we first obtain a relaxed convex problem of the original problem, which has a much smaller number of variables and can be solved efficiently. Then, we construct a near optimal solution of the original problem based on \textcolor{black}{an} optimal solution of the relaxed problem by connecting the construction to a packing problem.
This two-stage optimization method provides an upper bound on the performance gap between the optimal solution and the near optimal solution that can be easily evaluated.
In addition, the optimal solution of the relaxed problem reveals the impact of the caching gain of a layer \textcolor{black}{of} an SVC-based video or a description \textcolor{black}{of} a DASH-based video on its caching probability. 

\item Finally, \textcolor{black}{using numerical results based on real video sequences}, we show that for each type of videos, the proposed solution achieves a significant performance gain 
over baseline schemes \textcolor{black}{in the general and high user density regions}. In addition, we \textcolor{black}{demonstrate the respective operating regions of the proposed solutions for SVC-based and DASH-based videos}. 
\end{itemize}

\textcolor{black}{The key notations used in the paper are listed in Table~\ref{tab:para_system}}.
\begin{table}[h]
\caption{\textcolor{black}{Key notations.}}\label{tab:para_system}
\begin{center}
\vspace{-8mm}
\begin{scriptsize}
\textcolor{black}{
\begin{tabular}{|c!{\vrule width 1.5pt}c|}
\hline
Notation&Description\rule{0pt}{3mm}\\
\hline
$\Phi_h$, $\lambda_h$ & PPP for helpers, density of PPP for helpers\\
\hline
$P$, $W$, $\alpha$& transmit power at each helper, total bandwidth, path loss exponent\\
\hline
$N$, $\mathcal N$, $L$, $\mathcal L$ & number of videos, set of videos, number of video quality levels, set of video quality levels\\
\hline
$g\in\{{\rm SVC},{\rm DASH}\} $ & index for video encoding technique\\
\hline
$s_\ell$, $S_{{g},\ell}$ & size of layer $\ell$ of each SVC-based video, size of version $\ell$ of each $g$-based video\\
\hline
$a_n$, $b_{n,\ell}$  & probability of video $n$ being requested, probability of quality level for requested video $n$ being $\ell$\\
\hline
$C$, $\mathbf x$, $\mathcal X_{g}$  & cache size of each helper, cache content, cache content base for $g$-based videos\\
\hline
$p_{\mathbf x}$, $T_{n,\ell}$ & probability of cache content $\mathbf x$ being stored, probability of version $\ell$ of video $n$ being stored\\
\hline
$r_\ell$, $R_{{g},\ell}$ &  transmission rate for layer $\ell$ of each SVC-based video, transmission rate for version $\ell$ of each $g$-based video\\
\hline
$q_{g}(\mathbf p)$ & successful transmission probability for $g$-based videos\\
\hline
\end{tabular}}
\end{scriptsize}
\vspace{-12mm}
\end{center}
\end{table}

\section{System Model and Performance Metric}

\subsection{Network Model}\label{subsec:network model}


\begin{figure}[t]
\begin{center}
\subfigure[\small{SVC-based videos.}]
{\resizebox{6cm}{!}{\includegraphics{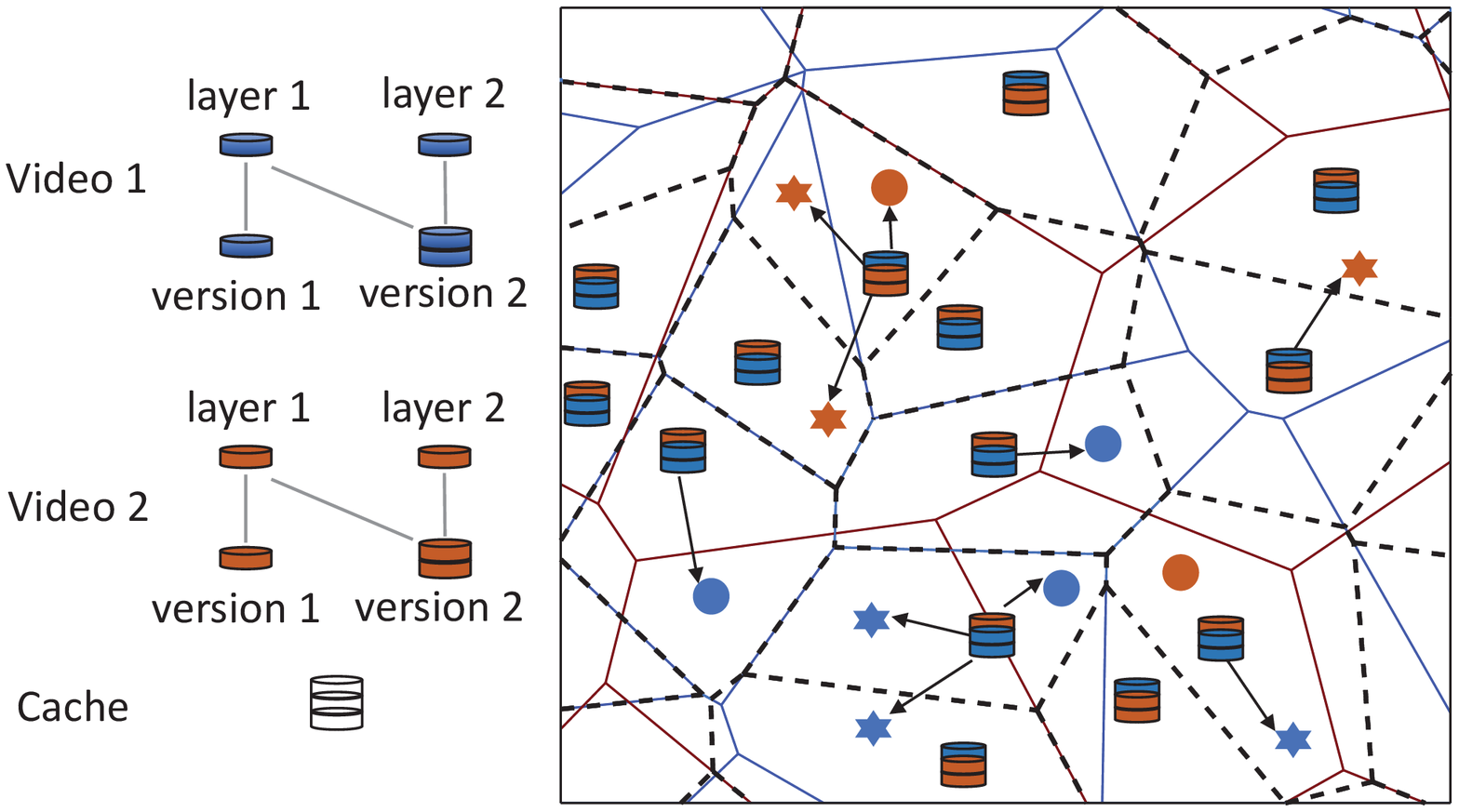}}}\hspace{10mm}
\subfigure[\small{DASH-based videos.}]
{\resizebox{7cm}{!}{\includegraphics{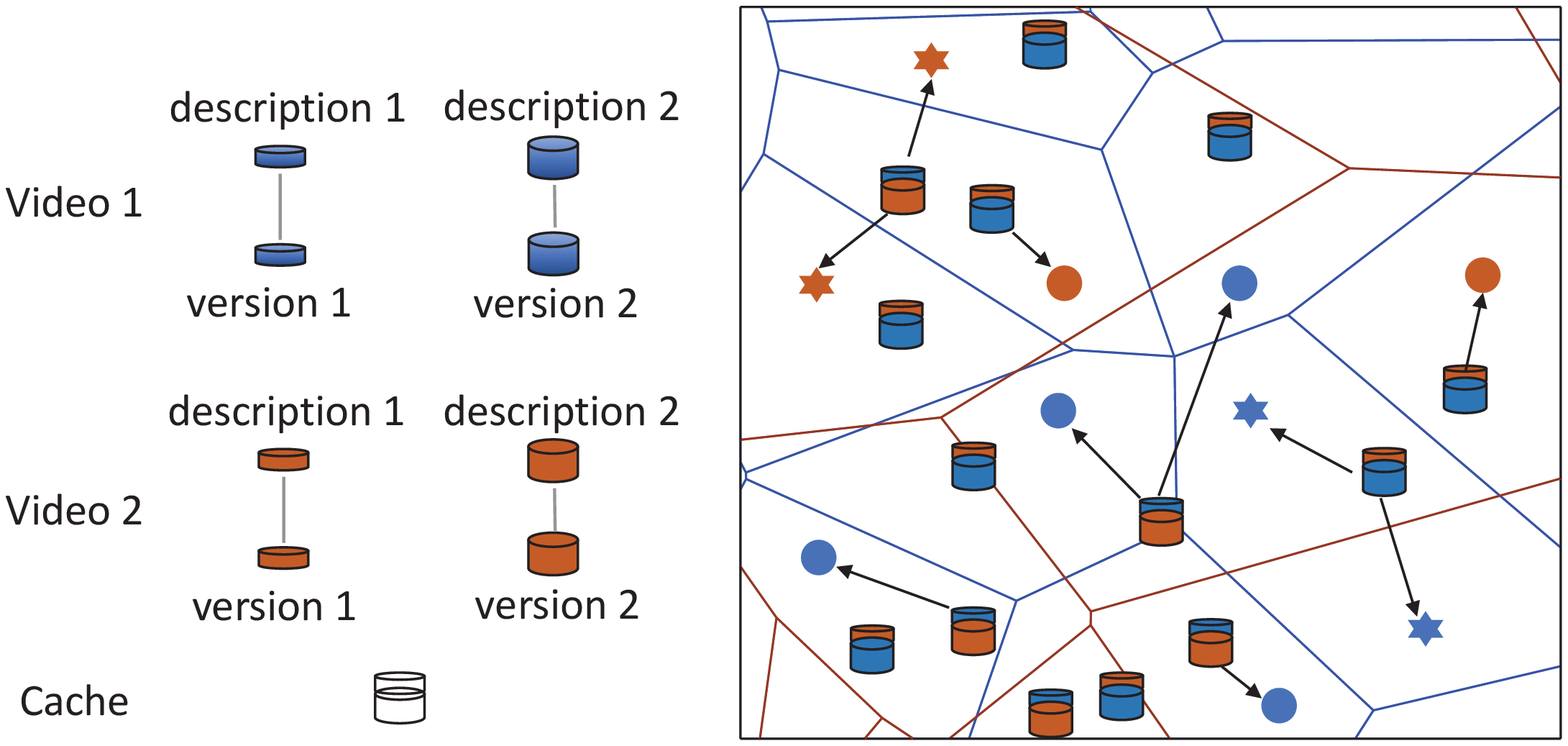}}}
\end{center}
\vspace{-4mm}
\caption{{\small System Model. $N=2$ and $L=2$. 
For SVC-based and DASH-based videos, each helper stores version $2$ of one video and version $1$ of the other.
Each circle (star) represents a user requesting a video with quality $1$ ($2$), the color of which indicates the video requested by the user.
In Fig.~\ref{fig:system_model}~(a), for each SVC-based video with quality $2$, there is a corresponding Voronoi tessellation represented by solid line segments in the same color as the video, which is determined by the locations of the helpers storing version $2$ of this video; for each SVC-based video with quality $1$, there is a corresponding Voronoi tessellation, which is determined by the locations of the helpers storing version $1$ or version $2$ of the video; the Voronoi tessellations for the two SVC-based videos with quality $1$ coincide, represented by black dash line segments.
In Fig.~\ref{fig:system_model}~(b), for each DASH-based video with quality $\ell\in\{1,2\}$, there is a corresponding Voronoi tessellation determined by the locations of the helpers storing version $\ell$ of this video; 
the Voronoi tessellations for video $1$ with quality $2$ and video $2$ with quality $1$ coincide, represented by blue solid line segments; the Voronoi tessellations for video $2$ with quality $2$ and video $1$ with quality $1$ coincide, represented by red solid line segments.
}}
\vspace{-8mm}
\label{fig:system_model}
\end{figure}

We consider a large-scale cache-enabled network, as shown in Fig.~\ref{fig:system_model}.  The locations of caching helpers are spatially distributed as a two-dimensional homogeneous Poisson point process (PPP)
$\Phi_h$ with density $\lambda_h$. The locations of users are distributed as an independent two-dimensional homogeneous PPP with density $\lambda_u$.
We consider downlink transmission. Each helper has one transmit antenna with transmission power $P$. Each user has one receive antenna. Consider a discrete-time narrow-band system of total bandwidth $W$ (in Hz) with time being slotted. At each slot,
all helpers transmit over the same frequency band. 
Consider both path loss and small-scale fading. Due to path loss, a transmitted signal with distance $d$ is attenuated by a factor $d^{-\alpha}$, where $\alpha>2$ is the path loss exponent. For small-scale fading, we assume Rayleigh fading channels, i.e., each small-scale channel at one slot over the whole frequency band $h\sim\mathcal {CN}(0, 1)$.

Let $\mathcal N\triangleq \{1,2,\cdots, N\}$ denote the set of $N$ videos in the network. Each video is encoded into $L$ versions of $L$ quality levels, 
and the $\ell$-th version provides the $\ell$-th highest quality, 
where $\ell\in\mathcal L\triangleq \{1,2,\cdots, L\}$. 
To achieve this, we consider 
two video encoding techniques.
\begin{itemize}
\item{\em SVC-based Videos}:
Consider encoding each video using SVC. In particular,
each video $n\in\mathcal N$ is encoded into $L$ layers. 
The successful decoding of layer $\ell$ requires the successful decodings of layers $1,\cdots, \ell-1$ 
for all $\ell \in \mathcal L\setminus\{1\}$, and \textcolor{black}{the successful decoding of} the base layer (i.e., layer $1$) does not rely on \textcolor{black}{that of any other layer}~\cite{Schwarz07}.
For ease of exposition, we also refer to layers $1,\cdots, \ell$ of an SVC-based video as the $\ell$-th version.
For \textcolor{black}{ease of analysis}, we assume that 
the size of layer $\ell$ of each SVC-based video is $s_\ell$ (in bits).
Then, the size of version $\ell$ of each SVC-based video is $S_{{\rm SVC},\ell}\triangleq \sum_{j=1}^\ell s_{j}$. 
For ease of exposition, we \textcolor{black}{let} $S_{{\rm SVC},0}=0$.

\item{\em DASH-based Videos}:
Consider encoding each video using HEVC or H.264 as in DASH. In particular, each video $n\in\mathcal N$ is encoded into $L$ descriptions~\cite{Hartanto2006}. For ease of exposition, we also refer to the $\ell$-th description of a DASH-based video as the $\ell$-th version.
For \textcolor{black}{ease of analysis}, we assume that the size of version $\ell$ of each DASH-based video is $S_{{\rm DASH},\ell}$ (in bits). 
\end{itemize}
For $g\in\{{\rm SVC},{\rm DASH}\}$, $S_{g,\ell}$ increases with $\ell$, i.e., $S_{g,1}<\cdots<S_{g,L}$. Note that on the one hand, due to the layered-encoding overhead, SVC uses more bits than the encoding in DASH to achieve the same quality level~\cite{Poularakis14INFOCOM}. On the other hand, due to the layered structure, SVC can provide more quality levels than the encoding in DASH, given the same number of bits. 

In this paper, we focus on one slot.
Each user randomly requests one video, which is video $n$ with probability $a_{n}\in[0,1]$, where $\sum_{n\in\mathcal N}a_{n}=1$.  
In addition, for each user requesting video $n$, the requested quality level is $\ell$ with probability $b_{n,\ell}\in[0,1]$, where $\sum_{\ell\in\mathcal L}b_{n,\ell}=1$. Note that $a_nb_{n,\ell}$ represents the popularity of video $n$ with quality $\ell$. The popularity distribution of the $N$ \textcolor{black}{multi-quality} 
videos is specified by $\mathbf a\triangleq (a_{n})_{n\in\mathcal N}$ and ${\mathbf b}_n\triangleq (b_{n,\ell})_{\ell\in\mathcal L}$, $n\in\mathcal N$. 
\textcolor{black}{Note that popularity distribution evolves at a slow timescale (e.g., on a daily or weekly basis), and can be estimated to certain extent using standard learning methodologies. As in \cite{Zhang17GLOBECOM}, \cite{Wu17Lett} and \cite{Choi18JSAC}, we assume that the popularity distribution is perfectly known to obtain first-order design insights}.\footnote{\textcolor{black}{The optimization results obtained in this paper can be extended to the case of imperfect popularity distribution using a robust optimization framework, and to the case of unknown popularity distribution using a stochastic optimization framework~\cite{YeTWC18}}.}

\subsection{Random Caching}

The network consists of cache-enabled helpers. In particular, each helper is equipped with a cache of size $\textcolor{black}{C}$ (in bits). We consider random caching for  SVC-based videos and DASH-based videos, respectively, as illustrated in Fig.~\ref{fig:system_model}.
We consider limited storage capacity by assuming $\textcolor{black}{C}<NS_{{\rm SVC},L}$ for SVC-based videos and $\textcolor{black}{C}< N\sum_{\ell=1}^LS_{{\rm DASH},\ell}$ for DASH-based videos, respectively.
For implementation simplicity in practice, assume that every user is served by only one helper.
Therefore, \textcolor{black}{given the layered structure of SVC-based videos}, basic cache components should be the $NL$ versions of the $N$ videos \textcolor{black}{for both SVC-based and DASH-based videos}. 
Each helper can store one cache content, which consists of basic cache components and is denoted by $\mathbf x\triangleq(x_{n,\ell})_{n\in\mathcal N,\ell\in\mathcal L}\in\{0,1\}^{N\times L}$. Here, $x_{n,\ell}=1$ indicates that cache content $\mathbf x$ contains version $\ell$ of video $n$, and $x_{n,\ell}=0$ otherwise. In the following, we specify cache contents for SVC-based videos and DASH-based videos, respectively.\footnote{\textcolor{black}{The purpose of imposing the constrains in \eqref{eqn:x_n_l_larger_K} and \eqref{eqn:x_n_l_larger_K_dash} is to reduce the computational complexity for solving the optimization problems which will be introduced shortly.}} 
\begin{itemize}
	\item{\em Cache Contents for SVC-based Videos}:
	For SVC-based videos,
we require: 
\vspace{-2mm}
\begin{align}
&\sum_{\ell\in\mathcal L}x_{n,\ell}\leq 1, \quad n\in\mathcal N,\label{eqn:x_n}\\
&\sum_{n\in\mathcal N} \sum_{\ell\in\mathcal L}x_{n,\ell}S_{{\rm SVC},\ell}\leq \textcolor{black}{C},\label{eqn:x_n_l_K}\\
&\sum_{n\in\mathcal N}\sum_{\ell\in\mathcal L}x_{n,\ell}' S_{{\rm SVC},\ell} > \textcolor{black}{C}, \quad \ \mathbf x'\in\{0,1\}^{N\times L}, \mathbf x' \mathbf 1_{L\times 1}\preceq \mathbf  1_{N\times 1}, \mathbf x'\mathbf z\succeq \mathbf x\mathbf z, \mathbf x'\neq \mathbf x.\label{eqn:x_n_l_larger_K}
\end{align}
Here, $\mathbf 1_{k\times 1}$ denotes a $k\times 1$ vector with all elements being 1, $\mathbf z\triangleq (1,2,\cdots, L)^T$, and $\succeq$ represents element-wise $\geq$. The constraints in \eqref{eqn:x_n} guarantee that each cache content $\mathbf x$  contains at most one version of an SVC-based video. The constraint in \eqref{eqn:x_n_l_K}  guarantees that the size of each cache content $\mathbf x$ is no greater than the cache size.
The constraints in \eqref{eqn:x_n_l_larger_K} guarantee that each cache content $\mathbf x$ is dominant in storage utilization. Let $\mathcal X_{\rm SVC}\triangleq \{\mathbf x\in\{0,1\}^{N\times L}:\eqref{eqn:x_n},\eqref{eqn:x_n_l_K},\eqref{eqn:x_n_l_larger_K}\}$ denote the cache content base 
for SVC-based videos.

\item{\em Cache Contents for DASH-based Videos}:
For DASH-based videos, we require: 
\begin{align}
&\sum_{n\in\mathcal N}\sum_{\ell\in\mathcal L} x_{n,\ell}S_{{\rm DASH},\ell}\leq \textcolor{black}{C},\label{eqn:x_n_l_K_dash}\\
&\sum_{n\in\mathcal N}\sum_{\ell\in\mathcal L} x_{n,\ell}' S_{{\rm DASH},\ell}> \textcolor{black}{C}, \quad \ \mathbf x'\in\{0,1\}^{N\times L}, \mathbf x'\succeq \mathbf x,\mathbf x'\neq \mathbf x.\label{eqn:x_n_l_larger_K_dash}
\end{align}
Note that different from SVC-based videos, we allow storing multiple versions of a DASH-based video at each helper. The constraints in \eqref{eqn:x_n_l_K_dash} and  \eqref{eqn:x_n_l_larger_K_dash} for DASH-based videos are comparable to those in \eqref{eqn:x_n_l_K} and  \eqref{eqn:x_n_l_larger_K} for SVC-based videos.
Let $\mathcal X_{\rm DASH}\triangleq \{\mathbf x\in\{0,1\}^{N\times L}:\eqref{eqn:x_n_l_K_dash},\eqref{eqn:x_n_l_larger_K_dash}\}$ denote the cache content base 
for DASH-based videos.

\end{itemize}

To provide spatial content diversity, for $g\in\{{\rm SVC},{\rm DASH}\}$, 
each helper randomly stores one cache content from cache content base $\mathcal X_g$, 
which is cache content $\mathbf x$ with probability $p_{\mathbf x}$ satisfying
\vspace{-2mm}
\begin{align}
&0\leq p_{\mathbf x}\leq 1, \quad \mathbf x\in\mathcal X_g,\label{eqn:constraint_0_1}\\
&\sum_{\mathbf x\in\mathcal X_g}p_{\mathbf x}=1.\label{eqn:constraint_sum}
\end{align}
Denote $\mathbf p\triangleq (p_{\mathbf x})_{{\mathbf x}\in\mathcal X_g}$.
To facilitate the analysis and optimization in later sections, based on $\mathbf p$, we also define the probability that a helper stores version $\ell$ of video $n$, i.e., 
\vspace{-2mm}
\begin{align}
T_{n,\ell}\triangleq \sum_{{\mathbf x}\in\mathcal X_g:x_{n,\ell}=1}p_{\mathbf x},\quad n\in\mathcal N, \ell\in\mathcal L.\label{eqn:svc_relation_T_p}
\end{align}
Denote $\mathbf T\triangleq(T_{n,\ell})_{n\in\mathcal N,\ell\in\mathcal L}$.
Note that the proposed random caching designs for SVC-based videos and DASH-based videos are different from the random caching design for independent single-quality files \textcolor{black}{with} the same file size in~\cite{Cui16TWC}.
In this paper, we focus on serving cached versions of $N$ videos at helpers to get first-order insights into cache-enabled wireless networks~\cite{Cui16TWC,Jiang17TWC}.

\subsection{Multicasting}
First, we introduce content-based user association under random caching for SVC-based videos and DASH-based videos, respectively.
\begin{itemize}
	\item{\em User Association under Random Caching for SVC-based Videos}:
Under random caching for SVC-based videos, a helper storing any version in $\{\ell,\ell+1,\cdots,L\}$ of SVC-based video $n\in\mathcal N$ can provide video $n$ with quality $\ell$. Thus, as illustrated in Fig.~\ref{fig:system_model}~(a), each user requesting video $n$ with quality $\ell$ is associated with (served by) the nearest helper storing cache content $\mathbf x\in\{\mathbf x\in\mathcal X_{\rm SVC}:\sum_{j=\ell}^L x_{n,j}=1\}$ containing any version in $\{\ell,\ell+1,\cdots,L\}$ 
of SVC-based video $n$. 

\item{\em User Association under Random Caching for DASH-based Videos}:
Under random caching for DASH-based videos, a helper storing only version $\ell$ of DASH-based video $n\in\mathcal N$ can provide video $n$ with quality $\ell$. Thus, as illustrated in Fig.~\ref{fig:system_model}~(b), each user requesting video $n$ with quality $\ell$ is associated with (served by) the nearest helper storing cache content $\mathbf x\in\{\mathbf x\in\mathcal X_{\rm DASH}:x_{n,\ell}=1\}$ containing version $\ell$ of DASH-based video $n$.
\end{itemize}
Note that the serving helper 
can offer the maximum long-term average receive power. 
The serving helper of a user may not be its geographically nearest helper and is statistically determined by the popularity distribution $\mathbf a$ and $\mathbf b_n$, $n\in\mathcal N$ of the $N$ videos as well as the caching distribution $\mathbf p$. 
Note that the user association for SVC-based videos is different from the one for single-quality files in~\cite{Cui16TWC} 
due to the layered structure of SVC;
the user association for DASH-based videos can be seen as an extension of the one for single-quality files in~\cite{Cui16TWC}, by treating $NL$ versions of $N$ DASH-based videos as $NL$ independent single-quality files with different file sizes.

Next, we consider multi-quality multicast for efficient dissemination of SVC-based videos and DASH-based videos, respectively. Consider a helper storing cache content $\mathbf x$.
Let $y_{n,\ell}\in\{0,1\}$ denote whether video $n$ with quality $\ell$ is requested by at least one of its associated users, where $y_{n,\ell}=1$ indicates that video $n$ with quality $\ell$ is requested and $y_{n,\ell}=0$ otherwise.
\begin{itemize}
	\item{\em Multicasting for SVC-based Videos}:
	Let $v_n\triangleq \max\{\ell y_{n,\ell}:\ell\in\mathcal L\}$ denote the highest quality level of video $n$ requested by its associated users, where $v_n=0$ represents that there is no request for video $n$.
If $v_n>0$, the helper transmits the first $v_n$ layers of SVC-based video $n$ only once to concurrently satisfy all received requests for SVC-based video $n$; if $v_n=0$, the helper does not transmit any layer of SVC-based video $n$.
Thus, $\mathbf v\triangleq (v_{n})_{n\in\mathcal N}$ can represent the transmitted layers of SVC-based videos. The helper transmits the $\sum_{n\in\mathcal N}v_n$ layers 
using frequency division multiple access (FDMA). In addition, to \textcolor{black}{utilize the layered structure of SVC-based videos for improving transmission efficiency}, 
the helper adopts proportional bandwidth allocation, i.e.,
the bandwidth allocated for transmitting a layer is proportional to its layer size. More specifically, for each SVC-based video $n\in\mathcal N$ with $v_n\geq 1$, the helper transmits layer $i\in\{1,\cdots,v_n\}$ over $\frac{s_i}{K_{\rm SVC}}$ of total bandwidth $W$ at rate $r_{i}$, where $K_{\rm SVC}\triangleq \sum_{n\in\mathcal N}S_{{\rm SVC},v_n}$ represents the total amount of data transmitted by the helper, referred to as the traffic load. 
We assume that the transmission rates of all layers, $r_{\ell}$, $\ell\in\mathcal L$, are proportional to their layer sizes, $s_\ell$, $\ell\in\mathcal L$, i.e., $\frac{r_{1}}{s_1}=\cdots=\frac{r_{L}}{s_L}$. Let $R_{\rm SVC,\ell}\triangleq \sum_{i=1}^\ell r_i$ denote the transmission rate for version $\ell$ of each SVC-based video. Note that $\frac{R_{{\rm SVC},1}}{S_{{\rm SVC},1}}=\cdots=\frac{R_{{\rm SVC},L}}{S_{{\rm SVC},L}}$.

\item{\em Multicasting for DASH-based Videos}:
If $y_{n,\ell}=1$, the helper transmits description $\ell$ of DASH-based video $n$ to satisfy received requests for DASH-based video $n$ with quality $\ell$; if $y_{n,\ell}=0$, the helper does not transmit description $\ell$ of DASH-based video $n$.
Thus, $\mathbf y\triangleq (y_{n,\ell})_{n\in\mathcal N,\ell\in\mathcal L}$ can represent the transmitted descriptions of DASH-based videos.
Similarly, the helper transmits the $\sum_{n\in\mathcal N}\sum_{\ell\in\mathcal L}y_{n,\ell}$ descriptions 
using FDMA, and
adopts proportional bandwidth allocation, i.e.,
the bandwidth allocated for transmitting a description is proportional to its description size. More specifically,
for each DASH-based video $n\in\mathcal N$ with $\sum_{\ell\in\mathcal L}y_{n,\ell}\geq 1$, the helper transmits description $i\in\{\ell\in\mathcal L:y_{n,\ell}=1\}$ over $\frac{S_{{\rm DASH},i}}{K_{\rm DASH}}$ of total bandwidth $W$ at rate $R_{{\rm DASH},i}$, where $K_{\rm DASH}\triangleq \sum_{n\in\mathcal N}\sum_{\ell\in\mathcal L}y_{n,\ell}S_{{\rm DASH},\ell}$
represents the traffic load of the helper.
We assume that the transmission rates of all descriptions, $R_{{\rm DASH},\ell}$, $\ell\in\mathcal L$, are proportional to their description sizes, $S_{{\rm DASH},\ell}$, $\ell\in\mathcal L$, i.e., $\frac{R_{{\rm DASH},1}}{S_{{\rm DASH},1}}=\cdots=\frac{R_{{\rm DASH},L}}{S_{{\rm DASH},L}}$.
\end{itemize}

Note that, by avoiding transmitting one layer or one description of a requested video multiple times, 
multicasting improves the utilization efficiency of the wireless medium and reduces the load of the wireless links, compared to the traditional connection-based (unicast) transmission~\cite{Cui16TWC,Jiang17TWC,Wang18TCOM}.

\subsection{Performance Metric}

In this paper, according to Slivnyak's theorem, we focus on a typical user $u_0$, which we assume without loss of generality to be located at the origin.
We consider an interference-limited network and neglect the background thermal noise \cite{Wu17Lett}. For analytical tractability, as in \cite{Zhang17GLOBECOM}, \cite{Wu17Lett}, \cite{Cui16TWC} and \cite{AndrewsTWCOffloading14}, we assume that all helpers are active. This is close to the actual situation of a heavily loaded network and \textcolor{black}{yields} the worst-case interference strength for $u_0$.
Suppose $u_0$ requests video $n$ with quality $\ell$.
The received signal of $u_0$ 
is given by
\begin{align}
&y_{g,n,\ell,0} = d_{0}^{-\frac{\alpha}{2}}h_{0}\textcolor{black}{s}_{g,n,\ell,0}+\sum_{b\in\Phi_b\setminus \{B_{g,n,\ell,0}\}}d_{b}^{-\frac{\alpha}{2}}h_{b}\textcolor{black}{s}_b,\quad\quad  g\in\{{\rm SVC}, {\rm DASH}\},\notag
\end{align}
where $B_{g,n,\ell,0}$ is the serving helper of $u_0$, $d_{0}$ is the distance between $u_0$ and $B_{g,n,\ell,0}$, $h_{0}\dis\mathcal {CN}(0,1)$ is the small-scale channel between $u_0$ and $B_{g,n,\ell,0}$, $\textcolor{black}{s}_{g,n,\ell,0}$ is the transmit signal from $B_{g,n,\ell,0}$, $d_b$ is the distance between $u_0$ and helper $b$, $h_b\dis\mathcal {CN}(0,1)$ is the small-scale channel between $u_0$ and helper $b$, and $\textcolor{black}{s}_b$ is the transmit signal from helper $b$.
The signal-to-interference ratio (SIR) of $u_0$ is given by 
\begin{align}
&{\rm SIR}_{g,n,\ell}=\frac{d_{0}^{-\alpha}|h_{0}|^2}{\sum_{b\in\Phi_b\setminus\{B_{g,n,\ell,0}\}}d_{b}^{-\alpha}|h_{b}|^2},\quad \quad g\in\{{\rm SVC},{\rm DASH}\}.\notag
\end{align}

Let $K_{g,n,\ell}$ denote the traffic load of $B_{g,n,\ell,0}$. 
The channel capacity for transmitting version $\ell$ of video $n$ is given by
$C_{g,n,\ell}\triangleq \frac{S_{g,\ell}W}{K_{g,n,\ell}}\log_2(1+{\rm SIR}_{g,n,\ell})$. 
The dissemination of version $\ell$ of video $n$ at rate $R_{g,\ell}$ can be successfully decoded if $C_{g,n,\ell}\geq R_{g,\ell}$.\footnote{\textcolor{black}{To obtain first-order design insights, we adopt the information-theoretical approach and consider capacity achieving code with arbitrarily small decoding error, as in~\cite{Wu17Lett,Zhang17GLOBECOM,Choi18JSAC}. Hence error propagation in decoding SVC-based videos is ignored. In addition, the proportional bandwidth allocation guarantees that the transmissions of all requested layers of an SVC-based video succeed or fail simultaneously.}} 
Then, the successful transmission probability of video $n$ with quality $\ell$ requested by $u_0$, 
denoted as $q_{g,n,\ell}(\mathbf p)$, 
is given by
\begin{align}
&q_{g,n,\ell}(\mathbf p)\triangleq \Pr\left[C_{g,n,\ell}\geq R_{g,\ell}\right]= \Pr\left[\frac{1}{K_{g,n,\ell}}\log_2(1+{\rm SIR}_{g,n,\ell})\geq \theta_g\right],\quad \quad g\in\{{\rm SVC},{\rm DASH}\},\notag
\end{align}
where the last equality is due to 
$\frac{R_{g,1}}{S_{g,1}W}=\cdots=\frac{R_{g,L}}{S_{g,L}W}\triangleq \theta_g$. 
Requesters are mostly concerned about whether their \textcolor{black}{requested} videos with desired qualities can be successfully received. Therefore, in this paper, we consider the successful transmission probability of a video with \textcolor{black}{a} certain quality randomly requested by $u_0$ as the network performance metric.
According to the total probability theorem,  the successful transmission probability of a video with \textcolor{black}{a} certain quality randomly requested by $u_0$, denoted as $q_{g}(\mathbf p)$, is given by 
\begin{align}
&q_{g}(\mathbf p) = \sum_{n\in\mathcal N}a_{n}\sum_{\ell\in\mathcal L} b_{n,\ell}q_{g,n,\ell}(\mathbf p), \quad \quad g\in\{{\rm SVC},{\rm DASH}\}.\label{eqn:STP_SVC_def}
\end{align}

\section{Performance Analysis and Optimization for SVC-based Videos}

\subsection{Performance Analysis for SVC-based Videos}
In this part, we analyze the successful transmission probability $q_{\rm SVC}(\mathbf p) $ for given caching distribution $\mathbf p$. In general, the traffic load $K_{{\rm SVC},n,\ell}$ and SIR ${\rm SIR}_{{\rm SVC},n,\ell}$ are correlated, as helpers with larger association regions have higher load and lower SIR (due to larger user to helper distances). However, the exact relationship between $K_{{\rm SVC},n,\ell}$ and ${\rm SIR}_{{\rm SVC},n,\ell}$ is very complex and is still not known. For analytical tractability, the dependence is ignored~\cite{Cui16TWC,AndrewsTWCOffloading14}. Then, from \eqref{eqn:STP_SVC_def}, we have $q_{\rm SVC}(\mathbf p)=\sum_{n\in\mathcal N}a_{n}\sum_{\ell\in\mathcal L} b_{n,\ell}\sum_{k\in\mathcal K_{{\rm SVC},n,\ell}}\Pr[K_{{\rm SVC},n,\ell}=k]
\Pr\left[\frac{1}{k}\log_2(1+{\rm SIR}_{{\rm SVC},n,\ell})\geq \theta_{\rm SVC}\right]$,
where $\mathcal K_{{\rm SVC},n,\ell}\triangleq \{\sum_{m\in\mathcal N}S_{{\rm SVC},v_m}: \mathbf v \preceq \mathbf x\mathbf z, v_n\geq \ell, \mathbf x\in\mathcal X_{\rm SVC}\}$.


First, we calculate the probability mass function (p.m.f.) of $K_{{\rm SVC},n,\ell}$. Note that 
at the serving helper, the number of cache components in the cache content and the number of layers \textcolor{black}{to be transmitted} for \textcolor{black}{a} cache  component are random, and different layers of an SVC-based video have different layer sizes, which make the calculation of $K_{{\rm SVC},n,\ell}$ very challenging.
For analytical tractability, adopting a commonly used approximation~\cite{Cui16TWC}, \cite{SGcellsize13}, 
we can calculate the p.m.f. of $K_{{\rm SVC},n,\ell}$ using tools from stochastic geometry. 

\begin{Lem}[p.m.f. of $K_{{\rm SVC},n,\ell}$]\label{Lem:pmf_K_svc_n_x}
The p.m.f. of $K_{{\rm SVC},n,\ell}$ is given by 
\vspace{-4mm}
\begin{align}
&\Pr\left[K_{{\rm SVC},n,\ell}=k\right]\notag\\
&= \sum\limits_{\mathbf x\in\mathcal X_{\rm SVC}:u_{n}(\mathbf x)\geq \ell}\frac{p_{\mathbf x}}{\sum_{j=\ell}^L{T_{n,j}}}\sum_{\mathbf v\in\mathcal {SQ}_{\mathbf x,n,\ell}(k)}\left(\mathbf 1[v_n> \ell]+\frac{\mathbf 1[v_n=\ell]\prod_{z=\ell+1}^{u_{n}(\mathbf x)}w_{n,z}}{W_{n,u_{n}(\mathbf x),v_n}}\right)\prod_{m\in\mathcal N:u_m(\mathbf x)>0}W_{m,u_m(\mathbf x),v_m},\notag
\end{align}
where $u_{n}(\mathbf x)\triangleq \sum_{\ell\in\mathcal L}\ell x_{n,\ell}$, 
$\mathcal {SQ}_{\mathbf x,n,\ell}(k) \triangleq \{\mathbf v:\sum_{m\in\mathcal N}S_{{\rm SVC},v_m}=k, v_m\in\{0,\cdots,u_{m}(\mathbf x)\},v_n\geq \ell\}$, 
$w_{n,\ell}=\left(1+\frac{a_{n}b_{n,\ell}\lambda_u}{3.5(\sum_{j=\ell}^LT_{n,j})\lambda_b}\right)^{-4.5}$, and
\begin{align}
&W_{m,j,i}
=\begin{cases}
(1-w_{m,i})\prod_{z=i+1}^jw_{m,z},\quad &i\in\{1,2,\cdots,j\}\\
\prod_{z=1}^jw_{m,z}, & i = 0
\end{cases}.\notag
\end{align}
\end{Lem}
\begin{IEEEproof}
Please refer to Appendix A.
\end{IEEEproof}

Note that $\sum_{j=\ell}^LT_{n,j}$ represents the probability that a helper stores any version in $\{\ell,\ell+1,\cdots,L\}$ 
of SVC-based video $n$,
and $w_{n,\ell}$ (which is a function of $\sum_{j=\ell}^LT_{n,j}$) 
represents the probability that a helper that stores any version in $\{\ell,\ell+1,\cdots,L\}$ of SVC-based video $n$ does not transmit layer $\ell$ of SVC-based video $n$. 
From Lemma~\ref{Lem:pmf_K_svc_n_x}, we can see that the physical layer parameters $\lambda_b$ and $\lambda_u$, the popularity distribution $\mathbf a$ and $\mathbf b_n$, $n\in\mathcal N$, and the caching distribution $\mathbf p$ jointly affect the p.m.f. of $K_{{\rm SVC},n,\ell}$. 



Next, we calculate the complementary cumulative distribution function (c.c.d.f.) of ${\rm SIR}_{{\rm SVC},n,\ell}$. 
Note that 
there are two types of interferers, namely, i) interfering helpers storing 
any version in $\{\ell,\ell+1,\cdots,L\}$ of SVC-based video $n$ (these helpers are further than the serving helper), and ii) interfering helpers storing 
any version in $\{1,2,\cdots,\ell-1\}$ of SVC-based video $n$ or not storing any version of SVC-based video $n$ (these helpers could be closer to $u_0$ than the serving helper). 
By carefully handling these two types of interferers, we can calculate the c.c.d.f. of ${\rm SIR}_{{\rm SVC},n,\ell}$ using tools from stochastic geometry. 
\begin{Lem}[c.c.d.f. of SIR]\label{Lem:distribution_SIR}
The c.c.d.f. of ${\rm SIR}_{{\rm SVC},n,\ell}$ is given by $\Pr[{\rm SIR}_{{\rm SVC},n,\ell}\geq\tau]=f\left(\tau,\sum_{j=\ell}^L T_{n,j}\right)$,
where
\begin{align}
&f(\tau,x)= \frac{x}{\textcolor{black}{D}_2(\tau)+\textcolor{black}{D}_1(\tau)x}.\label{eqn:f}
\end{align}
Here, $\textcolor{black}{D}_1(\tau) = 1+\frac{2}{\alpha}\tau^{\frac{2}{\alpha}}B'\left(\frac{2}{\alpha},1-\frac{2}{\alpha},\frac{1}{1+\tau}\right)-\frac{2}{\alpha}\tau^{\frac{2}{\alpha}}B\left(\frac{2}{\alpha},1-\frac{2}{\alpha}\right)$, $\textcolor{black}{D}_2(\tau) = \frac{2}{\alpha}\tau^{\frac{2}{\alpha}}B\left(\frac{2}{\alpha},1-\frac{2}{\alpha}\right)$, $B(a,b)\triangleq\int_{0}^{1}u^{a-1}(1-u)^{b-1}{\rm d}u$ denotes the beta function, and $B^{'}(a,b,z)\triangleq\int_{z}^{1}u^{a-1}(1-u)^{b-1}{\rm d}u$ ($0<z<1$) denotes the complementary incomplete beta function.
\end{Lem}
\begin{IEEEproof}
Please refer to Appendix B.
\end{IEEEproof}

From Lemma~\ref{Lem:distribution_SIR}, we can see that the impact of the physical layer parameters $\alpha$ and $\tau$ (captured by \textcolor{black}{$D_1(\tau) $ and $D_2(\tau)$}) and the impact of the caching distribution $\mathbf p$ on the c.c.d.f. of ${\rm SIR}_{{\rm SVC},n,\ell}$ are separated.
In addition, $\Pr[{\rm SIR}_{{\rm SVC},n,\ell}\geq\tau]$ depends on the probability that a helper stores any version in $\{\ell,\ell+1,\cdots,L\}$ of SVC-based video $n$, i.e., $\sum_{j=\ell}^LT_{n,j}$.


Based on Lemma~\ref{Lem:pmf_K_svc_n_x} and Lemma~\ref{Lem:distribution_SIR}, we can obtain 
$q_{\rm SVC}(\mathbf p)$ as follows.
\begin{Thm}[Performance for SVC-based Videos]\label{Thm:STP_SVC}
The successful transmission probability for SVC-based videos is given by
\begin{align}
q_{\rm SVC}(\mathbf p) = \sum_{n\in\mathcal N}a_{n}\sum_{\ell\in\mathcal L} b_{n,\ell}\sum_{k\in\mathcal K_{{\rm SVC},n,\ell}}\Pr[K_{{\rm SVC},n,\ell}=k]\Pr[{\rm SIR}_{{\rm SVC},n,\ell}\geq\tau_k],\notag
\end{align}
where $\Pr[K_{{\rm SVC},n,\ell}=k]$ is given by Lemma~\ref{Lem:pmf_K_svc_n_x}, $\Pr[{\rm SIR}_{{\rm SVC},n,\ell}\geq\tau_k]$ is given by Lemma~\ref{Lem:distribution_SIR}, and $\tau_k\triangleq 2^{k\theta_{\rm SVC} }-1$.
\end{Thm}


\begin{figure}[t]
\begin{center}
\subfigure[\small{SVC-based videos.}]
{\resizebox{5.2cm}{!}{\includegraphics{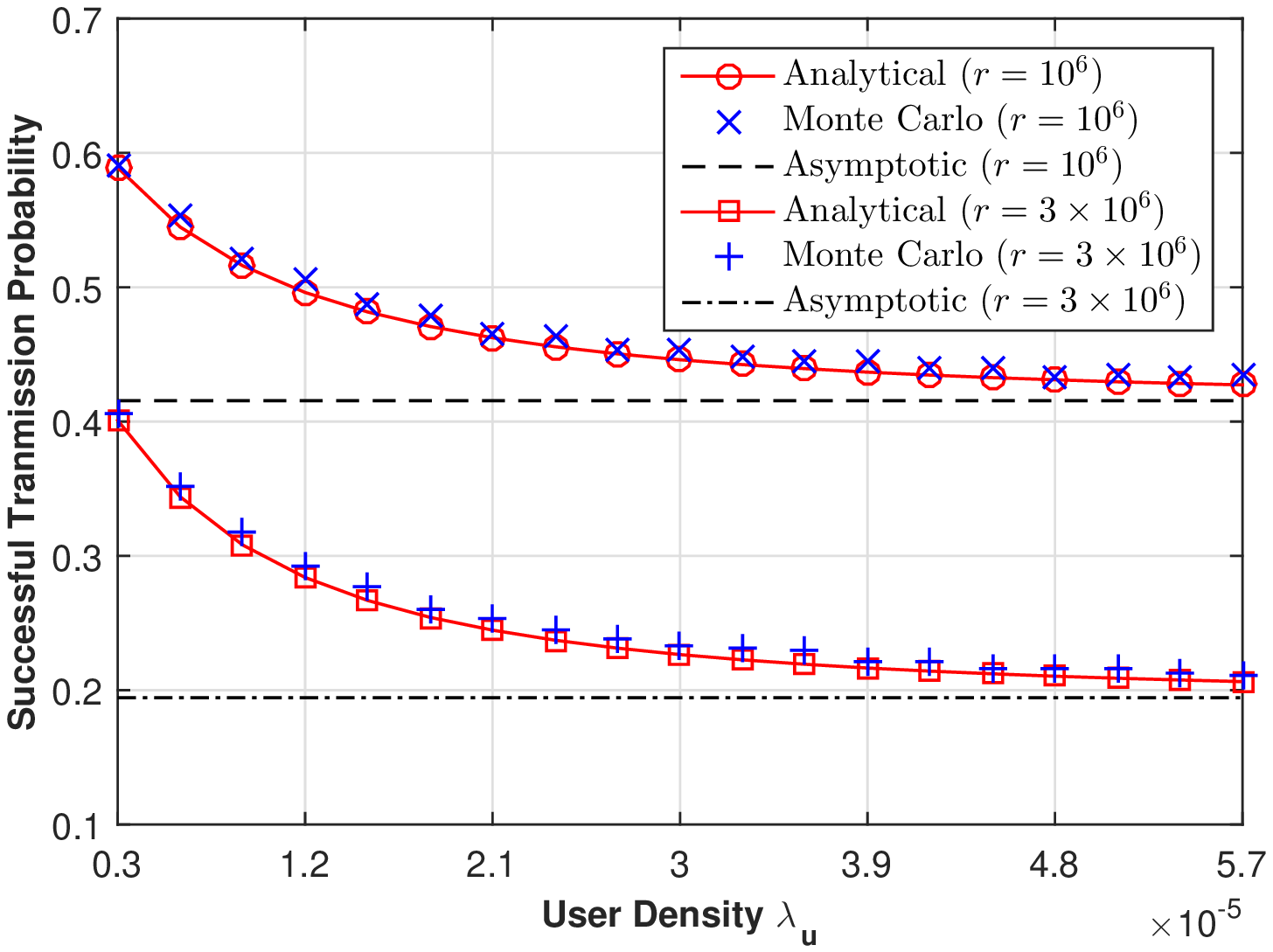}}}\hspace{10mm}
\subfigure[\small{DASH-based videos.}]
{\resizebox{5.2cm}{!}{\includegraphics{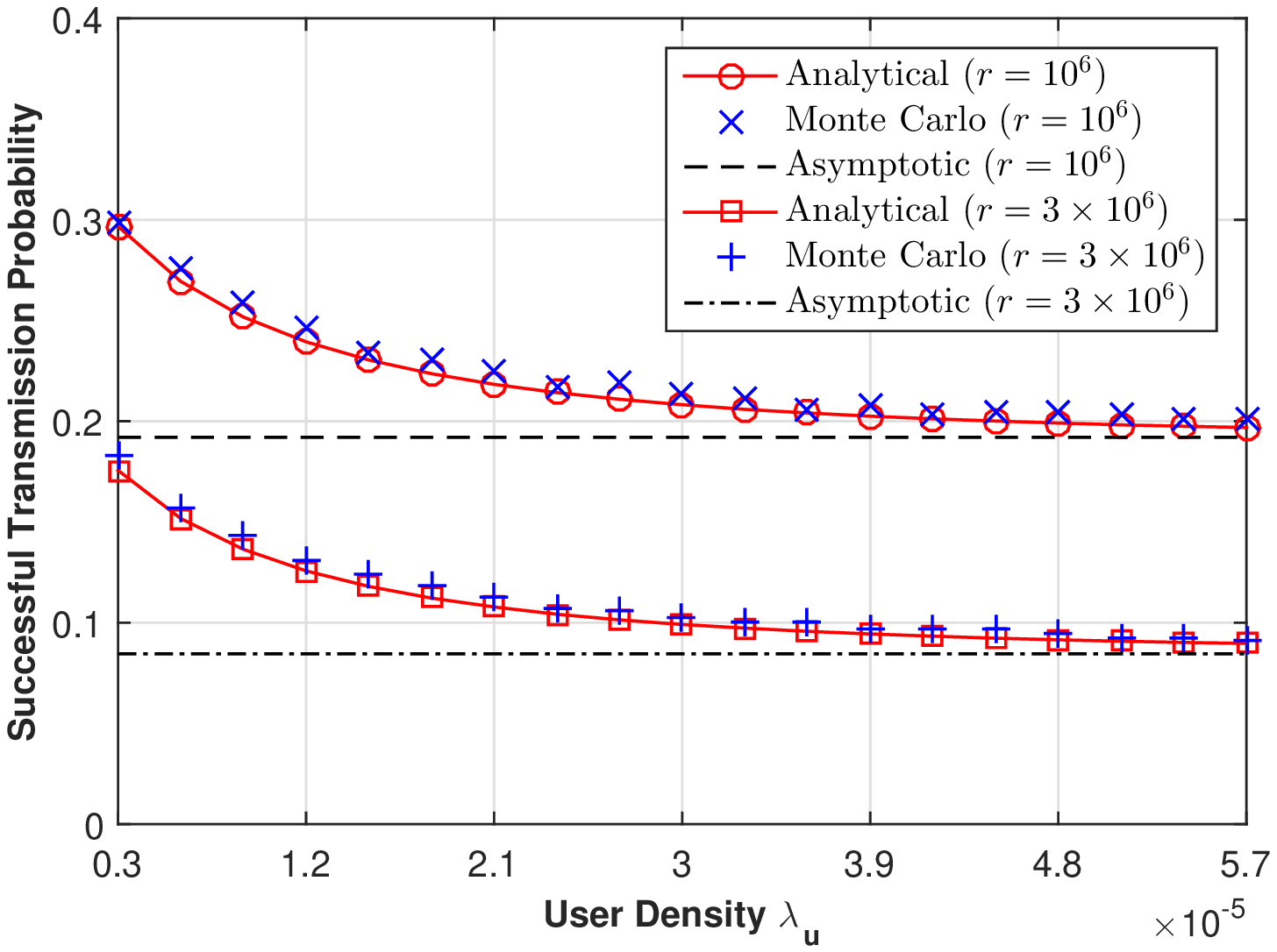}}}
\end{center}
\vspace{-2mm}
\caption{{\small Successful transmission probabilities for SVC-based videos and DASH-based videos versus user density $\lambda_u$. $N=5$, $L=3$, $\textcolor{black}{C}=8$, $s_\ell=1$, $r_{\ell}=r$, $\ell\in\mathcal L$, $S_{{\rm DASH,1}}=1$, $S_{{\rm DASH,2}}=2$, $S_{{\rm DASH,3}}=3$, $R_{{\rm DASH},1}=r$, $W=10\times10^6$, $\alpha = 4$,  $\lambda_b=3\times10^{-6}$,  $\mathbf x_1=(0,0,1;0,0,1;0,1,0;0,0,0;0,0,0)$, $\mathbf x_2=(0,0,1;0,0,1;0,0,0;0,1,0;0,0,0)$, $\mathbf x_3=(0,0,1;0,0,0;0,0,1;0,0,0;0,1,0)$, $\mathbf x_4=(0,1,0;0,1,0;1,0,0;0,0,1;0,0,0)$, $\mathbf x_5=(1,0,0;0,1,0;0,1,0;1,0,0; 0,1,0)$, $\mathbf p=(0.2,0.2,0.2,0.2,0.2)$, $b_{n,\ell}=\frac{1}{L}$ and  $a_{n}=\frac{n^{-\gamma}}{\sum_{n\in\mathcal N}n^{-\gamma}}$ with $\gamma=1$. \textcolor{black}{Here, we consider a simple illustration example with parameters chosen according to \cite{Poularakis16INFOCOM, Zhang17WCNC,Poularakis14INFOCOM}, to verify the analytical results for SVC-based and DASH-based videos.}}}
\vspace{-4mm}
\label{fig:SVC_anal_vs_MC_vs_asymp}
\end{figure}


To obtain design insights into caching and multicasting for SVC-based videos, we analyze the successful transmission probability in the high user density region. Note that the high user density region, where the gain of multicast over unicast achieves the maximum, is widely considered when studying multicast performance~\cite{Cui16TWC,Wang18TCOM}.
When $\textcolor{black}{C}\gg S_{{\rm SVC},L}$, by \eqref{eqn:x_n_l_larger_K}, we know that $\frac{\textcolor{black}{C}-\sum_{n\in\mathcal N}\sum_{\ell\in\mathcal L}x_{n,\ell}S_{{\rm SVC},\ell}}{\textcolor{black}{C}}\approx 0$ for all $\mathbf x\in\mathcal X_{\rm SVC}$. 
Approximating $\sum_{n\in\mathcal N}\sum_{\ell\in\mathcal L}x_{n,\ell}S_{{\rm SVC},\ell}$ with $\textcolor{black}{C}$,
from Theorem~\ref{Thm:STP_SVC}, we have the following corollary.
\begin{Cor}[Performance for SVC-based Videos when $\lambda_u\to \infty$]\label{Cor:STP_SVC_asymp}
\begin{align}
\lim_{\lambda_u\to\infty}  q_{\rm SVC}(\mathbf p)&= \sum_{n\in\mathcal N}\sum_{\ell\in\mathcal L} a_{n}b_{n,\ell}f(\tau_{\textcolor{black}{C}},\sum_{j=\ell}^LT_{n,j})\triangleq q_{{\rm SVC},\infty}(\mathbf T),\notag
\end{align}
where 
$f(\tau,x)$ is given by \eqref{eqn:f}. 
\end{Cor}

From Corollary~\ref{Cor:STP_SVC_asymp}, we can see that the successful transmission probability of version $\ell$ of SVC-based video $n$ in the high user density region $f(\tau_{\textcolor{black}{C}},\sum_{j=\ell}^LT_{n,j})$ is a concave increasing function of $\sum_{j=\ell}^LT_{n,j}$. In addition, $q_{{\rm SVC},\infty}(\mathbf T)$ is a concave increasing function of $\mathbf T$, since $\frac{x}{D_2(\tau_{\textcolor{black}{C}})+D_1(\tau_{\textcolor{black}{C}})x}$ is a concave function of $x$, a nonnegative weighted sum of concave functions is concave, and composition with a linear function preserves concavity.
Fig.~\ref{fig:SVC_anal_vs_MC_vs_asymp}~(a) plots the successful transmission probability for SVC-based videos versus the user density $\lambda_u$. From Fig.~\ref{fig:SVC_anal_vs_MC_vs_asymp}~(a), we can see that each ``Analytical'' curve (plotted using Theorem~\ref{Thm:STP_SVC}) closely matches the corresponding ``Monte Carlo'' curve, verifying Theorem~\ref{Thm:STP_SVC} and demonstrating the accuracy of the approximations adopted; when $\lambda_u$ increases, the gap between each ``Analytical'' curve and the corresponding ``Asymptotic'' curve (plotted using Corollary~\ref{Cor:STP_SVC_asymp}) decreases, verifying Corollary~\ref{Cor:STP_SVC_asymp}.

\subsection{Performance Optimization  for SVC-based Videos}\label{subsec:opt_SVC}

In this part, we optimize the caching distribution $\mathbf p$ to maximize the successful transmission probability $q_{{\rm SVC},\infty}(\mathbf T)$ in the high user density region.
\textcolor{black}{Numerical results will show that the obtained solution also achieves promising performance in the general user density region}.

\subsubsection{Problem Formulation}
We would like to maximize $q_{{\rm SVC},\infty}(\mathbf T)$ by carefully optimizing $\mathbf p$ under the constraints in \eqref{eqn:constraint_0_1} and \eqref{eqn:constraint_sum}.\footnote{\textcolor{black}{There are several formulations with different objective functions and constraints, each with its own meaning or application scenario. In this paper, we choose to consider the basic formulations with only continuous variables which correspond to the case of hard quality requirements. The proposed solutions can be extended to other formulations with both continuous variables and discrete variables (representing selection of quality levels for requested videos)}.} 

\begin{Prob}[Random Caching and Multicasting for SVC-based Videos]\label{Prob:SVC}
\vspace{-2mm}
\begin{align}
q_{\rm SVC}^*\triangleq \max_{\mathbf p}  &  \quad  q_{{\rm SVC},\infty}(\mathbf T)\notag\\
s.t. \ & \quad \eqref{eqn:constraint_0_1}, \eqref{eqn:constraint_sum}.\notag
\end{align}
\end{Prob}

The objective function $q_{{\rm SVC},\infty}(\mathbf T)$ of Problem~\ref{Prob:SVC} is a concave function of $\mathbf p$, since $q_{{\rm SVC},\infty}(\mathbf T)$ is a concave function of $\mathbf T$, and $\mathbf T$ is a linear function of $\mathbf p$ (note that composition of a concave function with a linear function preserves concavity).
In addition, noting that the equality and inequality constraint functions are linear, Problem~\ref{Prob:SVC} is a convex problem.
To solve Problem~\ref{Prob:SVC}, we need to first construct all the cache contents in the cache content base $\mathcal X_{\rm SVC}$, the number of which increases exponentially with $N$ and $\textcolor{black}{C}$. 
\textcolor{black}{The number of optimization variables of Problem~\ref{Prob:SVC} is the cardinality of $\mathcal X_{\rm SVC}$, which is exceedingly large for large $N$ and $\textcolor{black}{C}$}. Thus, Problem~\ref{Prob:SVC} cannot be solved with acceptable \textcolor{black}{computational} complexity for a practical network where $N$ and $\textcolor{black}{C}$ are usually very large.

\subsubsection{Near Optimal Solution}

In this part, we propose a two-stage optimization method to obtain a low-complexity near optimal solution of Problem~\ref{Prob:SVC}. 
In Stage I, we formulate and solve a relaxed problem of Problem~\ref{Prob:SVC} with $\mathbf T$ as the optimization variable instead of $\mathbf p$. Specifically, by~\eqref{eqn:x_n_l_K}, \eqref{eqn:constraint_0_1}, \eqref{eqn:constraint_sum} and \eqref{eqn:svc_relation_T_p}, we have
\vspace{-2mm}
\begin{align}
&T_{n,\ell}=\sum_{{\mathbf x}\in\mathcal X_{\rm SVC}:x_{n,\ell}=1}p_{\mathbf x}\geq 0,\quad n\in\mathcal N, \ell\in\mathcal L,\label{eqn:svc_constraint_positive}\\
&\sum_{\ell\in\mathcal L} T_{n,\ell}=\sum_{\mathbf x\in\mathcal X_{\rm SVC}:\sum_{\ell\in\mathcal L}x_{n,\ell}= 1}p_{\mathbf x} \leq  \sum_{\mathbf x\in\mathcal X_{\rm SVC}}p_{\mathbf x}=1,\quad n\in\mathcal N,\label{eqn:svc_constraint_n}\\
&\sum_{n\in\mathcal N}\sum_{\ell\in\mathcal L}S_{{\rm SVC},\ell} T_{n,\ell}=\sum_{n\in\mathcal N}\sum_{\ell\in\mathcal L}S_{{\rm SVC},\ell} \sum_{\mathbf x\in\mathcal X_{\rm SVC}:x_{n,\ell}=1}p_{\mathbf x}=\sum_{\mathbf x\in\mathcal X_{\rm SVC}}p_{\mathbf x}\sum_{n\in\mathcal N}\sum_{\ell\in\mathcal L}S_{{\rm SVC},\ell}x_{n,\ell}\leq \textcolor{black}{C}.\label{eqn:svc_constraint_cache_size}
\end{align}
Note that the constraints in~\eqref{eqn:svc_constraint_n} guarantee that the probability of a helper storing any version of video $n$ cannot exceed $1$.  The constraint in~\eqref{eqn:svc_constraint_cache_size} guarantees that the average occupied storage at each helper cannot exceed the cache size $\textcolor{black}{C}$.
Therefore, Problem~\ref{Prob:SVC} can be relaxed to \textcolor{black}{the following one}.
\begin{Prob}[Relaxed Problem of Problem~\ref{Prob:SVC}]\label{Prob:SVC_relaxed}
\vspace{-2mm}
\begin{align}
q^*_{{\rm SVC},{\rm ub}}\triangleq \max_{\mathbf T}  &  \quad  q_{{\rm SVC},\infty}(\mathbf T) \notag\\
s.t. \ & \quad \eqref{eqn:svc_constraint_positive},\eqref{eqn:svc_constraint_n},\eqref{eqn:svc_constraint_cache_size}.\notag
\end{align}
Let $\mathbf T_{\rm SVC}^*\triangleq (T_{{\rm SVC},n,\ell}^*)_{n\in\mathcal N,\ell\in\mathcal L}$ denote an optimal solution of Problem~\ref{Prob:SVC_relaxed}.
\end{Prob}

Given a feasible solution $\mathbf p$ of Problem~\ref{Prob:SVC}, we can find a corresponding feasible solution $\mathbf T$ of Problem~\ref{Prob:SVC_relaxed} using~\eqref{eqn:svc_relation_T_p}.
Given a feasible solution $\mathbf T$ of Problem~\ref{Prob:SVC_relaxed}, there may not exist a feasible solution $\mathbf p$  of Problem~\ref{Prob:SVC} satisfying~\eqref{eqn:svc_relation_T_p}.
The optimal value of Problem~\ref{Prob:SVC_relaxed} can serve as an upper bound for that of Problem~\ref{Prob:SVC}, i.e., $q_{\rm SVC}^*\leq q^*_{{\rm SVC},{\rm ub}}$.
Similar to Problem~\ref{Prob:SVC}, Problem~\ref{Prob:SVC_relaxed} is a convex problem. The number of optimization variables in Problem~\ref{Prob:SVC_relaxed} is $NL$ which is much smaller than that in Problem~\ref{Prob:SVC}, facilitating the optimization when $N$ and $\textcolor{black}{C}$ are large. We can obtain an optimal solution $\mathbf T_{\rm SVC}^*$ of Problem~\ref{Prob:SVC_relaxed} using any off-the-shelf interior-point solver (e.g., CVX).

Note that Slater's condition is satisfied for Problem~\ref{Prob:SVC_relaxed}, implying that strong duality holds. Using KKT conditions, we can obtain a semi-closed-form solution of Problem~\ref{Prob:SVC_relaxed}, which will be used for analyzing optimality properties of Problem~\ref{Prob:SVC_relaxed}. 

\begin{Lem}[Optimal Solution of Problem~\ref{Prob:SVC_relaxed}]\label{Lem:SVC_closed_form_optimal}
The optimal solution of Problem~\ref{Prob:SVC_relaxed} is given by 
\begin{align}
&T_{{\rm SVC},n,\ell}^*=\begin{cases}
\frac{1}{D_1(\tau_C)}\left(\sqrt{\frac{a_{n}b_{n,1}D_2(\tau_C)}{v^*s_1-\lambda_{n,1}^*+\eta_n^*}}-\sqrt{\frac{a_{n}b_{n,2}D_2(\tau_C)}{v^*s_2-\lambda_{n,2}^*+\lambda_{n,1}^*}}\right), &\quad \ell=1\\
\frac{1}{D_1(\tau_C)}\left(\sqrt{\frac{a_{n}b_{n,\ell}D_2(\tau_C)}{v^*s_\ell-\lambda_{n,\ell}^*+\lambda_{n,{\ell-1}}^*}}-\sqrt{\frac{a_{n}b_{n,{\ell+1}}D_2(\tau_C)}{v^*s_{\ell+1}-\lambda_{n,{\ell+1}}^*+\lambda_{n,{\ell}}^*}}\right), &\quad \ell\in\{2,\cdots,L-1\}\\
\frac{1}{D_1(\tau_C)}\sqrt{\frac{a_{n}b_{n,L}D_2(\tau_C)}{v^*s_L-\lambda_{n,L}^*+\lambda_{n,{L-1}}^*}}-\frac{D_2(\tau_C)}{D_1(\tau_C)},  &\quad  \ell=L
\end{cases}, \notag
\end{align}
where $T_{{\rm SVC},n,\ell}^*$, $\lambda_{n,\ell}^*\geq 0$, $\eta_n^*\geq 0$ and $v^*\geq 0$ satisfy $\lambda_{n,\ell}^*T_{{\rm SVC},n,\ell}^*=0$ for all $n\in\mathcal N$ and $\ell\in\mathcal L$, $\eta_n^*(1-\sum_{\ell \in\mathcal L}T^*_{{\rm SVC},n,\ell})=0$ for $n\in\mathcal N$, and $v^*(\textcolor{black}{C}-\sum_{n\in\mathcal N}\sum_{\ell\in\mathcal L}S_{{\rm SVC},\ell} T^*_{{\rm SVC},n,\ell})=0$.
\end{Lem}

Based on Lemma~\ref{Lem:SVC_closed_form_optimal}, we have the following result.
\begin{Lem}[Optimality Properties of Problem~\ref{Prob:SVC_relaxed}]\label{Lem:svc_property_opt_T}
(i) For all $n\in\mathcal N$ and $\ell\in\{1,2,\cdots,L-1\}$, if $\frac{b_{n,\ell}}{s_\ell}\leq \frac{b_{n,\ell+1}}{s_{\ell+1}}$, then $T_{{\rm SVC},n,\ell}^*=0$.
(ii) For all $n\in\mathcal N$ with $T_{{\rm SVC},n,j}^*>0$, $j\in\mathcal L$ and $\ell\in\{1,\cdots,L-2\}$, if $\sqrt{\frac{b_{n,\ell+1}}{s_{\ell+1}}}-\sqrt{\frac{b_{n,\ell+2}}{s_{\ell+2}}}\geq \sqrt{\frac{b_{n,\ell}}{s_{\ell}}}-\sqrt{\frac{b_{n,\ell+1}}{s_{\ell+1}}}$, then $T_{{\rm SVC},n,\ell+1}^*\geq T_{{\rm SVC},n,\ell}^*$.
(iii) For all $n_1,n_2\in\mathcal N$ with $T_{{\rm SVC},n_1,j}^*,T_{{\rm SVC},n_2,j}^*>0$, $j\in\mathcal L$ and for all $\ell\in\mathcal L$, if $a_{n_1}b_{n_1,\ell}\geq a_{n_2}b_{n_2,\ell}$, then $\sum_{j=\ell}^LT_{{\rm SVC},n_1,j}^*\geq \sum_{j=\ell}^LT_{{\rm SVC},n_2,j}^*$.
\end{Lem}
\begin{IEEEproof}
Please refer to Appendix C.
\end{IEEEproof}

Note that $\frac{b_{n,\ell}}{s_\ell}=\frac{b_{n,\ell}}{S_{{\rm SVC},\ell}-S_{{\rm SVC},\ell-1}}$ can be interpreted as the caching gain of layer $\ell$ \textcolor{black}{of} SVC-based video $n$. 
Property (i) indicates that version $\ell$ of SVC-based video $n$ will not be stored if $\frac{b_{n,\ell}}{s_\ell}\leq \frac{b_{n,\ell+1}}{s_{\ell+1}}$. This is because storing  version $\ell+1$ of SVC-based video $n$ can satisfy more requests per unit storage than storing version $\ell$ of SVC-based video $n$. 
In addition, from Property (i), we can conclude that if $\frac{b_{n,1}}{s_1}\leq \cdots\leq \frac{b_{n,L}}{s_L}$, only version $L$ of SVC-based video $n$ will be stored. 
Property (ii) indicates that the probability of storing version $\ell+1$ of SVC-based video $n$ is no smaller than that of storing version $\ell$ of SVC-based video $n$, if $\sqrt{\frac{b_{n,\ell+1}}{s_{\ell+1}}}-\sqrt{\frac{b_{n,\ell+2}}{s_{\ell+2}}}\geq \sqrt{\frac{b_{n,\ell}}{s_{\ell}}}-\sqrt{\frac{b_{n,\ell+1}}{s_{\ell+1}}}$. 
Property (iii) indicates that the probability of storing any version in $\{\ell,\ell+1,\cdots,L\}$ of SVC-based video $n_1$ is no smaller than that of storing any version in $\{\ell,\ell+1,\cdots,L\}$ of SVC-based video $n_2$, if $a_{n_1}b_{n_1,\ell}\geq a_{n_2}b_{n_2,\ell}$. This is because storing any version in $\{\ell,\ell+1,\cdots,L\}$ of SVC-based video $n_1$ can satisfy more requests with the same storage resource than storing any version in $\{\ell,\ell+1,\cdots,L\}$ of SVC-based video $n_2$. In addition, Property (iii) implies that if $a_{n_1}b_{n_1,\ell}= a_{n_2}b_{n_2,\ell}$, then $\sum_{j=\ell}^LT_{{\rm SVC},n_1,j}^*= \sum_{j=\ell}^LT_{{\rm SVC},n_2,j}^*$, as $\sum_{j=\ell}^LT_{{\rm SVC},n_1,j}^*= \sum_{j=\ell}^LT_{{\rm SVC},n_2,j}^*$ is equivalent to $\sum_{j=\ell}^LT_{{\rm SVC},n_1,j}^*\geq  \sum_{j=\ell}^LT_{{\rm SVC},n_2,j}^*$ and $\sum_{j=\ell}^LT_{{\rm SVC},n_1,j}^*\leq \sum_{j=\ell}^LT_{{\rm SVC},n_2,j}^*$.

\begin{figure}[t]
\begin{center}
\subfigure[\small{Algorithm~\ref{alg:SVC_opt} for} SVC-based videos.]
{\resizebox{7cm}{!}{\includegraphics{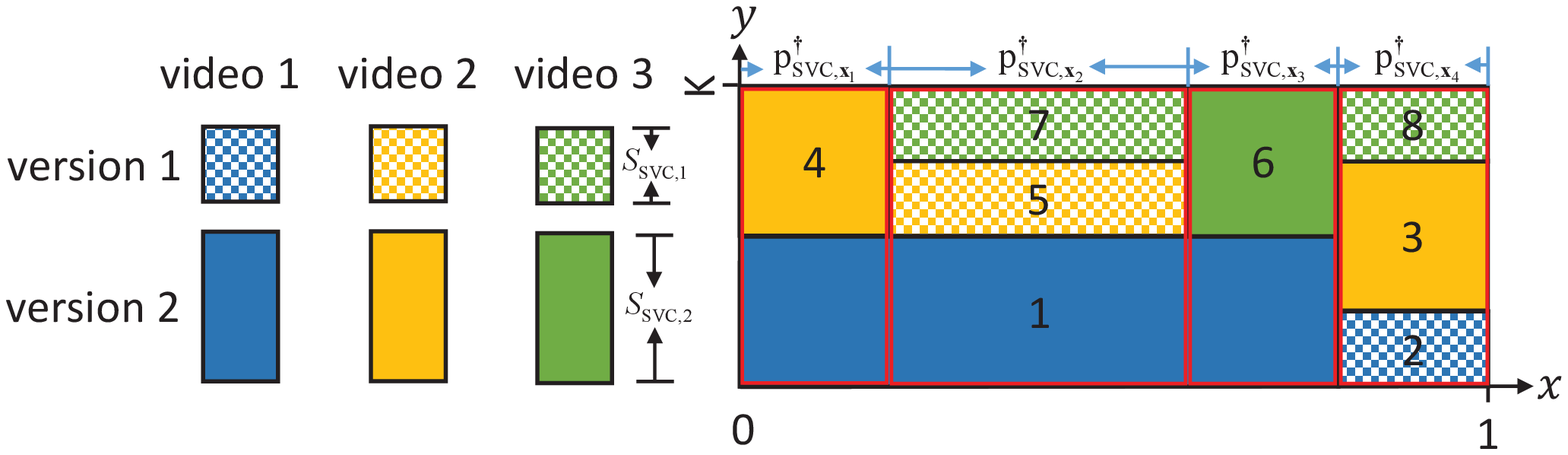}}}\hspace{10mm}
\subfigure[\small{Algorithm~\ref{alg:DASH_opt} for DASH-based videos.}]
{\resizebox{7cm}{!}{\includegraphics{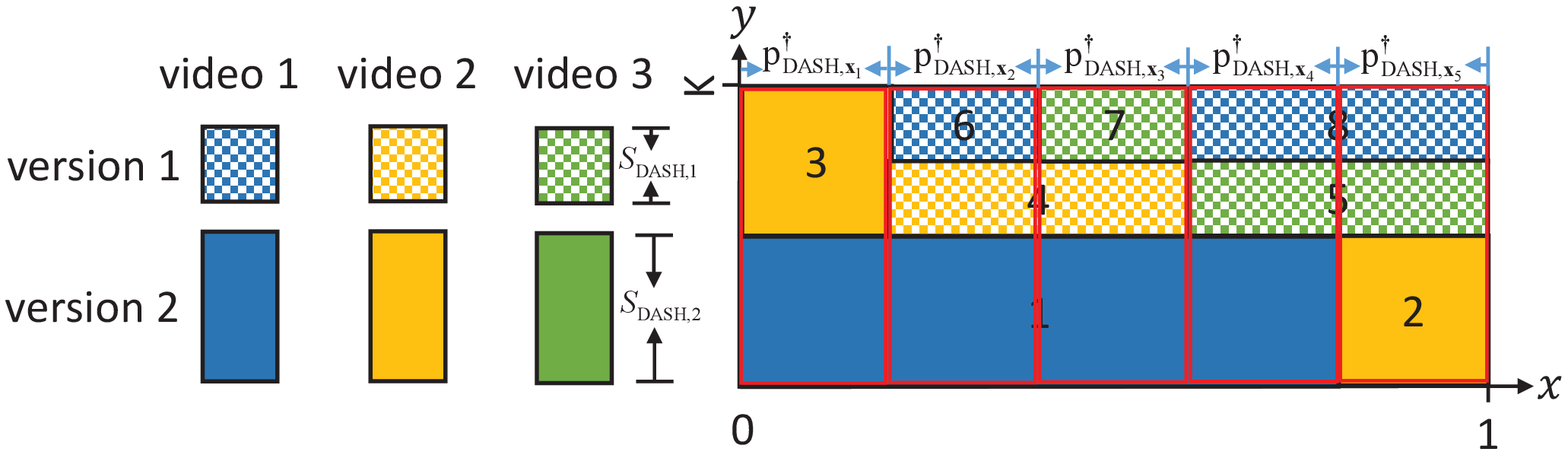}}}
\end{center}
\vspace{-4mm}
\caption{{\small Illustrations of Algorithm~\ref{alg:SVC_opt} and Algorithm~\ref{alg:DASH_opt}. $N=3$, $L=2$, $\textcolor{black}{C}=4$, $S_{{\rm SVC},1}=1$, $S_{{\rm SVC},2}=2$, $\mathbf T_{\rm SVC}^*=(0.2, 0.8;0.4, 0.4;0.4,0.3)$, $\sum_{\ell=1}^LS_{{\rm SVC},\ell} T^*_{{\rm SVC},1,\ell}>\sum_{\ell=1}^LS_{{\rm SVC},\ell} T^*_{{\rm SVC},2,\ell}>\sum_{\ell=1}^LS_{{\rm SVC},\ell} T^*_{{\rm SVC},3,\ell}$, $S_{{\rm DASH},1}=1$, $S_{{\rm DASH},2}=2$, $\mathbf T_{\rm DASH}^*=(0.2, 0.8;0.4, 0.4;0.4,0.3)$, $T^*_{{\rm DASH},1,2}\geq T^*_{{\rm DASH},2,2}\geq T^*_{{\rm DASH},2,1}\geq T^*_{{\rm DASH},3,1}\geq T^*_{{\rm DASH},3,2}\geq T^*_{{\rm DASH},1,1}$.
The numbers in the rectangles represent the placing order. }}
\vspace{-4mm}
\label{fig:form_combination}
\end{figure}

In Stage II, based on \textcolor{black}{an} optimal solution $\mathbf T_{{\rm SVC}}^*$ of Problem~\ref{Prob:SVC_relaxed}, we develop a low-complexity algorithm to construct a \textcolor{black}{caching} distribution which serves as a near optimal solution of Problem~\ref{Prob:SVC}, as summarized in Algorithm~\ref{alg:SVC_opt}. An illustration example is shown in Fig.~\ref{fig:form_combination}~(a).
The key idea  is as follows. A cache with size $\textcolor{black}{C}$ is represented by a rectangle box of length $1$ and height $\textcolor{black}{C}$, occupying region $[0,1]\times[0,\textcolor{black}{C}]$, and version $\ell$ of SVC-based video $n$ is represented by a rectangle of length $T_{{\rm SVC},n,\ell}^*$ and height $S_{{\rm SVC},\ell}$, which can be further partitioned vertically into multiple rectangles (for packing the rectangle box). We pack the rectangle box with rectangles representing video versions in a certain manner so that the projections of the rectangles for the same video on the $x$-axis do not overlap. The details are summarized in Steps $1-26$. It is clear that the resulting empty space of the rectangle box is smaller than $S_{{\rm SVC},L}$. Based on the packing of the rectangle box, we determine a \textcolor{black}{caching} distribution $\mathbf p_{\rm SVC}^{\dagger}\triangleq (p_{{\rm SVC},\mathbf x}^{\dagger})_{\mathbf x\in\mathcal X_{\rm SVC}}$ that satisfies the constraints in \eqref{eqn:constraint_0_1} and \eqref{eqn:constraint_sum} (i.e., is a feasible solution of Problem~\ref{Prob:SVC}). The details are summarized in Steps $27-34$. Note that we do not need to construct cache content base $\mathcal X_{\rm SVC}$ in Algorithm~\ref{alg:SVC_opt}. Instead, we only need to construct an effective cache content base $\widetilde{\mathcal X}_{\rm SVC}\subseteq \mathcal X_{\rm SVC}$, which is usually much smaller than $\mathcal X_{\rm SVC}$. \textcolor{black}{Note that when $L=1$, Algorithm~\ref{alg:SVC_opt} reduces to the probabilistic caching method in~\cite{ICC15Giovanidis}}.\footnote{\textcolor{black}{Algorithm~\ref{alg:SVC_opt} is a non-trivial extension of the probabilistic caching method in~\cite{ICC15Giovanidis}.}}

\begin{algorithm} \caption{Near Optimal Solution of Problem~\ref{Prob:SVC} ($g={\rm SVC}$)}
\small{\begin{algorithmic}[1]
\STATE Initialize $\mathcal A = [0,1]\times[0,\textcolor{black}{C}]$, $\mathcal I_n=[0,1]$ for all $n\in\mathcal N$, $\mathcal O_{n,\ell}=\emptyset$ for all $n\in\mathcal N$ and $\ell\in\mathcal L$,  $\mathcal P= \{0,1\}$, and $\mathbf R=\mathbf T_{g}^*$.
\STATE Sort  $\sum_{\ell\in\mathcal L}S_{{g},\ell} R_{n,\ell}$, $n\in\mathcal N$ in decreasing order. Let $\sum_{\ell\in\mathcal L}S_{{g},\ell} R_{(1),\ell}\geq \cdots \geq \sum_{\ell\in\mathcal L}S_{{g},\ell} R_{(N),\ell}$ be the resulting sequence.
\FOR{$n=1:N$}
\FOR{$\ell=L:-1:1$}
\WHILE{$R_{(n),\ell}>0$}
\STATE Set $y_b=\min\{y:(x,y)\in\mathcal A,x\in \mathcal I_{(n)}\}$.
\IF{$y_b\leq \textcolor{black}{C}-S_{{g},\ell}$}
\STATE Set $x_l=\min\{x\in \mathcal I_{(n)}: (x,y_b)\in\mathcal A\}$ and $x_r=\max\{x\in \mathcal I_{(n)}: a(x_l,y_b)+(1-a)(x,y_b)\in \mathcal A\ \text{for all}\ a\in[0,1]\}$.
\IF {$R_{(n),\ell}\leq x_r-x_l$}
\STATE Set $\mathcal I_{(n)} =\mathcal  I_{(n)}\setminus[x_l,x_l+R_{(n),\ell}]$, $\mathcal O_{(n),\ell}=\mathcal O_{(n),\ell}\cup [x_l,x_l+R_{(n),\ell}]$, $\mathcal A = \mathcal A\setminus ([x_l,x_l+R_{(n),\ell}]\times[y_b,y_b+S_{{g},\ell}])$, $\mathcal P= \mathcal P\cup \{x_l+R_{(n),\ell}\}$ and $R_{(n),\ell}=0$.
\ELSE
\STATE Set $\mathcal I_{(n)} = \mathcal I_{(n)}\setminus[x_l,x_r]$, $\mathcal O_{(n),\ell}=\mathcal O_{(n),\ell}\cup [x_l,x_r]$, $\mathcal A = \mathcal A\setminus ([x_l,x_r]\times[y_b,y_b+S_{{g},\ell}])$, $\mathcal P= \mathcal P\cup \{x_r\}$, and $R_{(n),\ell}=R_{(n),\ell}-(x_r-x_l)$.
\ENDIF
\ELSE
\STATE Set $R_{(n),\ell}=0$.
\ENDIF
\ENDWHILE
\ENDFOR
\ENDFOR
\STATE Initialize $T_{{g},n,\ell}^{\dagger}=\lambda(\mathcal O_{n,\ell})$ for all $n\in\mathcal N$ and $\ell\in\mathcal L$, where $\lambda(\mathcal O_{n,\ell})$ denotes the measure of the Lebesgue measurable set $\mathcal O_{n,\ell}$. Sort all $x\in\mathcal P$ in decreasing order. Let $x_{(1)}<x_{(2)}<\cdots<$ be the resulting sequence.
\FOR{$i=1:|\mathcal P|-1$}
\STATE Set $y_b=\min\{y:(x,y)\in\mathcal A,x\in [x_{(i)},x_{(i+1)}]\}$, $j=\max\{j\in \mathcal L\cup\{0\}: y_b+S_{{g},j}\leq \textcolor{black}{C}\}$.
\IF{$j\in\mathcal L$}
\STATE Set $m=\mathop{\arg\max}\limits_{n\in\mathcal N:[x_{(i)},x_{(i+1)}]\subseteq \mathcal I_n}a_{n}\sum\limits_{\ell=1}^j b_{n,\ell}(f(\tau_C,\sum_{z=\ell}^LT_{{g},n,z}^{\dagger}+x_{(i+1)}-x_{(i)})-f(\tau_C,\sum_{z=\ell}^LT_{{g},n,z}^{\dagger}))$, $\mathcal O_{m,j}=\mathcal O_{m,j}\cup [x_{(i)},x_{(i+1)}]$, and $T_{{g},m,j}^{\dagger}=T_{{g},m,j}^{\dagger}+x_{(i+1)}-x_{(i)}$.
\ENDIF
\ENDFOR
\STATE Initialize $\widetilde{\mathcal X}_{g}=\emptyset$.
\FOR{$i=1:|\mathcal P|-1$}
\STATE Set $\mathbf x=\mathbf 0$. 
\FOR{$(n,\ell)=\mathcal N\times \mathcal L$}
\STATE Set $x_{n,\ell}=\mathbf 1[[x_{(i)},x_{(i+1)}]\subseteq\mathcal O_{n,\ell}]$.
\ENDFOR
\STATE Set $p_{{g},\mathbf x}^{\dagger}=x_{(i+1)}-x_{(i)}$ and $\widetilde{\mathcal X}_{g}=\widetilde{\mathcal X}_{g}\cup\{\mathbf x\}$.
\ENDFOR
\end{algorithmic}}\normalsize\label{alg:SVC_opt}
\end{algorithm}

\textcolor{black}{Now, we characterize 
the distance between $\mathbf T_{\rm SVC}^*$ and $\mathbf T_{\rm SVC}^{\dagger}\triangleq(T_{{\rm SVC},n,\ell}^{\dagger})_{n\in\mathcal N,\ell\in\mathcal L}$ with $T_{{\rm SVC},n,\ell}^{\dagger}\triangleq \sum_{{\mathbf x}\in\mathcal X_{\rm SVC}:x_{n,\ell}=1}p_{{\rm SVC},\mathbf x}^{\dagger}$. 
As $\sum_{n\in\mathcal N}\sum_{\ell\in\mathcal L}R_{n,\ell}<1$ and $\max_i {x_{i}}-\min_i x_i\leq 1$, we have
$\parallel\mathbf T_{\rm SVC}^*-\mathbf T_{\rm SVC}^\dagger\parallel_1\leq\sum_{n\in\mathcal N}\sum_{\ell\in\mathcal L}R_{n,\ell}+\left(\max_i {x_{i}}-\min_i x_i\right)<2$. 
Note that $\parallel \mathbf T_{\rm SVC}^* \parallel_1\geq \frac{C}{S_{{\rm SVC},L}}$. Thus, we also have  
$\frac{\parallel \mathbf T_{\rm SVC}^*-\mathbf T_{\rm SVC}^\dagger\parallel_1}{\parallel \mathbf T_{\rm SVC}^* \parallel_1}< \frac{2S_{{\rm SVC},L}}{C}$,
which indicates that the relative difference between $\mathbf T_{\rm SVC}^*$ and $\mathbf T_{\rm SVC}^\dagger$ is negligible when $\textcolor{black}{C}\gg S_{{\rm SVC},L}$. Therefore, in general, $\mathbf p_{\rm SVC}^{\dagger}$ can serve as a near optimal solution of Problem~\ref{Prob:SVC}. In addition, we show that $\mathbf p_{\rm SVC}^{\dagger}$ is optimal in some special cases. When $L\geq 2$ and $\sum_{n\in\mathcal N}\sum_{\ell\in\mathcal L}R_{n,\ell}=0$ (i.e., the rectangle box right after Step 19 of Algorithm~\ref{alg:SVC_opt} is fully occupied), $\mathbf p_{\rm SVC}^{\dagger}$ and $\mathbf T_{\rm SVC}^*$ satisfy \eqref{eqn:svc_relation_T_p}. By noting that $q_{{\rm SVC},\infty}(\mathbf T_{\rm SVC}^*)\geq q_{\rm SVC}^*$,  
we know that in this case, $\mathbf p^{\dagger}_{\rm SVC}$ is an optimal solution of Problem~\ref{Prob:SVC}. When $L=1$, Algorithm~\ref{alg:SVC_opt} reduces to the probabilistic caching method in~\cite{ICC15Giovanidis}, indicating that $\mathbf p_{\rm SVC}^{\dagger}$ and $\mathbf T_{\rm SVC}^*$ satisfy \eqref{eqn:svc_relation_T_p}. Similarly,  we know that in this case, 
$\mathbf p^{\dagger}_{\rm SVC}$ is an optimal solution of Problem~\ref{Prob:SVC}}. 

\begin{figure}[t]
\begin{center}
\subfigure[\small{SVC-based videos.}]
{\resizebox{5.2cm}{!}{\includegraphics{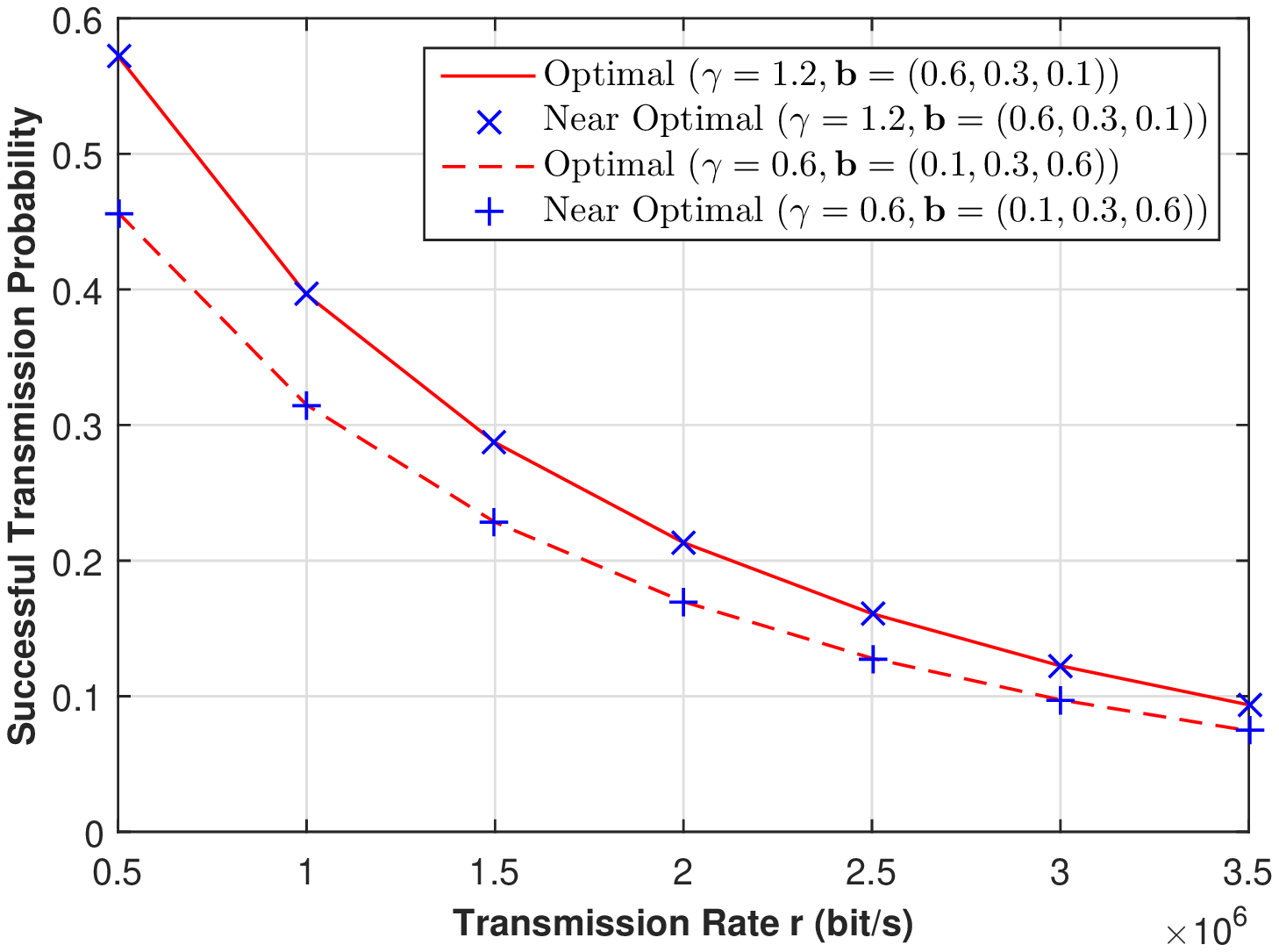}}}\hspace{10mm}
\subfigure[\small{DASH-based videos.}]
{\resizebox{5.2cm}{!}{\includegraphics{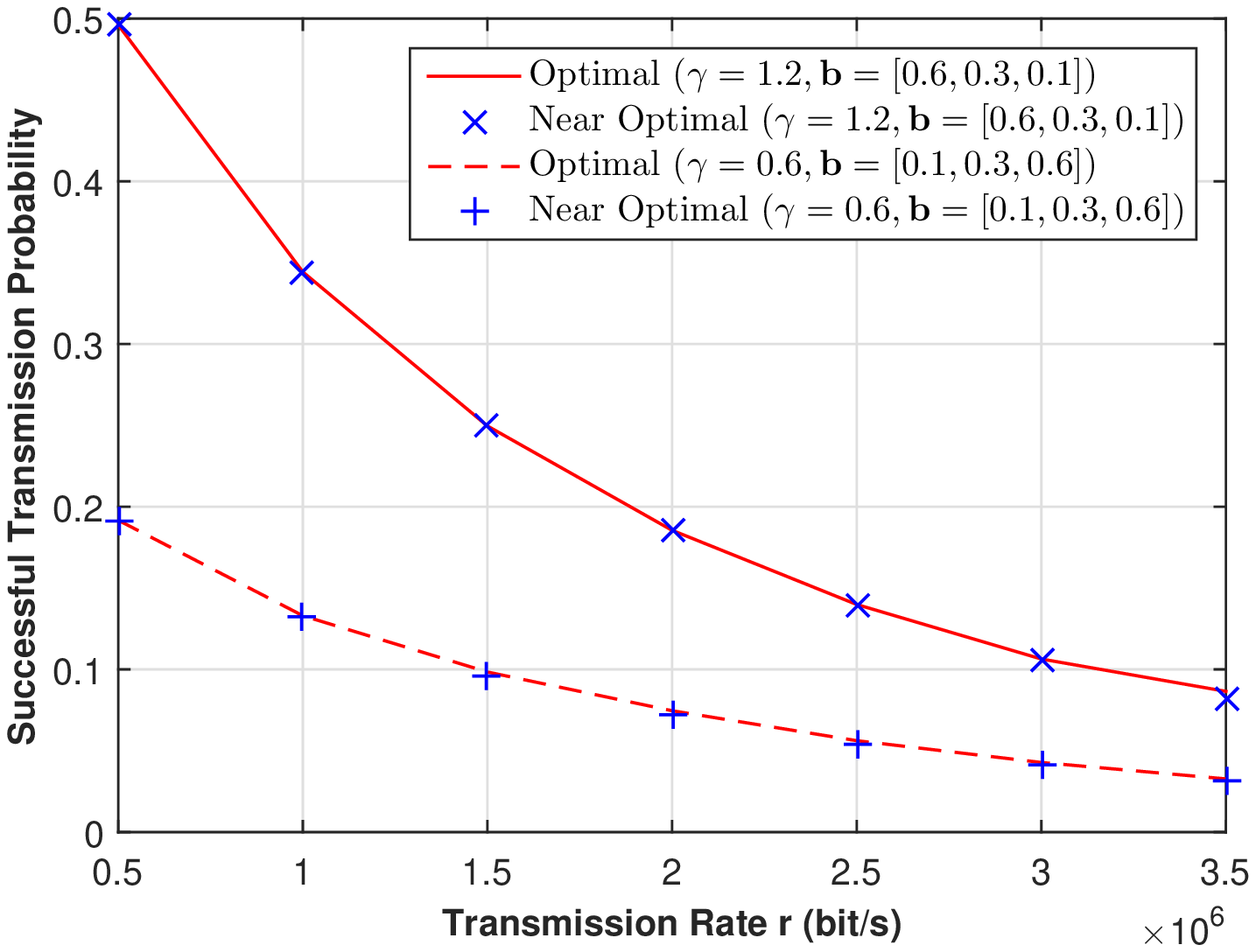}}}
\end{center}
\vspace{-2mm}
\caption{{\small Successful transmission probabilities for SVC-based videos and DASH-based videos versus transmission rate. $N=9$, $L=3$, $\textcolor{black}{C}=15$,  $s_\ell=1$, $r_{\ell}=r$, $\ell\in\mathcal L$, $S_{{\rm DASH},1}=1$, $S_{{\rm DASH},2}=2$, $S_{{\rm DASH},3}=3$, $R_{{\rm DASH},1}=r$, $W=10\times10^6$, $\alpha = 4$, $\mathbf b_n=\mathbf b$, $a_{n}=\frac{n^{-\gamma}}{\sum_{n\in\mathcal N}n^{-\gamma}}$. \textcolor{black}{Here, we consider a simple illustration example with parameters chosen according to \cite{Poularakis16INFOCOM, Zhang17WCNC,Poularakis14INFOCOM}, to verify the optimization results for SVC-based and DASH-based videos.}}}
\vspace{-4mm}
\label{fig:opt_ub_lb}
\end{figure}

Denote  $q_{\rm SVC}^{\dagger}\triangleq q_{{\rm SVC},\infty}(\mathbf T_{\rm SVC}^{\dagger})$. Since $q_{\rm SVC}^{\dagger} \leq q_{\rm SVC}^*\leq q^*_{{\rm SVC},{\rm ub}}$, we have $q_{\rm SVC}^*-q_{\rm SVC}^{\dagger}\leq q^*_{{\rm SVC},{\rm ub}} -  q_{\rm SVC}^{\dagger}$. Therefore, this two-stage optimization method also provides an upper bound on the performance gap between the optimal solution and the near optimal solution, i.e., $q^*_{{\rm SVC},ub} - q_{\rm SVC}^{\dagger}$, which can be easily evaluated.

In Fig.~\ref{fig:opt_ub_lb}~(a) and Table~\ref{tab:firsttable}, we use a numerical example to compare the optimal solution of Problem~\ref{Prob:SVC} obtained by CVX and the proposed near optimal solution obtained by the two-stage optimization method for SVC-based videos in both successful transmission probability and \textcolor{black}{computational} complexity.
We can see that the successful transmission probability of the proposed near optimal solution \textcolor{black}{is almost the same as} that of the optimal solution. In contrast, the average computation time for the optimal solution is $165$ times of that for the near optimal solution. These demonstrate the applicability and effectiveness of the near optimal solution.

\begin{table}
\caption{\textcolor{black}{Computational} complexities \textcolor{black}{of} the proposed solutions for SVC-based videos and DASH-based videos. The parameters are the same as those in Fig.~\ref{fig:opt_ub_lb}.}
\vspace{-4mm}
\centering
\small{\subtable[SVC-based videos]{
       \begin{tabular}{|c|c|c|}
       \hline
       & near optimal & optimal\\
       \hline
        Time & 1 & 165\\
        \hline
       \end{tabular}
       \label{tab:firsttable}
}
\hspace{10mm}
\subtable[DASH-based videos]{
       \begin{tabular}{|c|c|c|}
       \hline
       &near optimal & optimal\\
       \hline
       Time & 1 & 367\\
       \hline
       \end{tabular}
       \label{tab:secondtable}
}}\normalsize
\vspace{-8mm}
\end{table}

\section{Performance Analysis and Optimization for DASH-based Videos}

\subsection{Performance Analysis  for DASH-based Videos}

In this part, we analyze the successful transmission probability $q_{\rm DASH}(\mathbf p) $ for given caching distribution $\mathbf p$. 
Similarly, for analytical tractability, the dependence between the traffic load $K_{{\rm DASH},n,\ell}$ and SIR ${\rm SIR}_{{\rm DASH},n,\ell}$ is ignored. Then from \eqref{eqn:STP_SVC_def}, we have $q_{\rm DASH}(\mathbf p)=\sum_{n\in\mathcal N}a_{n}\sum_{\ell\in\mathcal L} b_{n,\ell}\sum_{k\in\mathcal K_{{\rm DASH},n,\ell}}\\ \Pr[K_{{\rm DASH},n,\ell}=k]
\Pr\left[\frac{1}{k}\log_2(1+{\rm SIR}_{{\rm DASH},n,\ell})\geq \theta_{\rm DASH}\right]$,
where $\mathcal K_{{\rm DASH},n,\ell}\triangleq \{\sum\limits_{m\in\mathcal N}\sum\limits_{j\in\mathcal L}y_{m,j}S_{{\rm DASH},j}: \mathbf y \preceq \mathbf x, y_{n,\ell}=1, \mathbf x\in\mathcal X_{\rm DASH}\}$.

First, we calculate the p.m.f. of $K_{{\rm DASH},n,\ell}$.
For analytical tractability, adopting a commonly used approximation, 
we can calculate the p.m.f. of $K_{{\rm DASH},n,\ell}$. 

\begin{Lem}[p.m.f. of $K_{{\rm DASH},n,\ell}$]\label{Lem:pmf_K_dash_n_x}
The p.m.f. of $K_{{\rm DASH},n,\ell}$ is given by
\begin{align}
&\Pr\left[K_{{\rm DASH},n,\ell}=k\right]\notag\\
&= \sum_{\mathbf x\in\mathcal X_{\rm DASH}:x_{n,\ell}=1}\frac{p_{\mathbf x}}{T_{n,\ell}}\sum_{\mathbf y\in\widetilde{\mathcal {SQ}}_{\mathbf x,n,\ell}(k)}\prod_{(m,j)\in \mathcal N\times \mathcal L:x_{m,j}=1,y_{m,j}=0}\widetilde{w}_{m,j}\prod_{(m,j)\in \mathcal N\times \mathcal L:x_{m,j}=1,y_{m,j}=1}(1-\widetilde{w}_{m,j}),\notag
\end{align}
where $\widetilde{\mathcal {SQ}}_{\mathbf x,n,\ell}(k)\triangleq \{\mathbf y:\sum_{m\in\mathcal N}\sum_{j\in\mathcal L}y_{m,j}S_{{\rm DASH},j}=k, \mathbf y\preceq \mathbf x,y_{n,\ell}=1\}$, and $\widetilde{w}_{n,\ell}=\left(1+\frac{a_{n}b_{n,\ell}\lambda_u}{3.5T_{n,\ell}\lambda_b}\right)^{-4.5}$.
\end{Lem}
\begin{IEEEproof}
Lemma~\ref{Lem:pmf_K_dash_n_x} can be proved in a similar way to Lemma~\ref{Lem:pmf_K_svc_n_x}. We omit the details due to page limitation.
\end{IEEEproof}

Note that $\widetilde{w}_{n,\ell}$ (which is a function of  $T_{n,\ell}$) 
represents the probability that a helper that stores version $\ell$ of DASH-based video $n$ does not transmit version $\ell$ of DASH-based video $n$. 
From Lemma~\ref{Lem:pmf_K_dash_n_x}, we can see that the physical layer parameters $\lambda_b$ and $\lambda_u$, the popularity distribution $\mathbf a$ and $\mathbf b_n$, $n\in\mathcal N$, and the caching distribution $\mathbf p$ jointly affect the p.m.f. of $K_{{\rm DASH},n,\ell}$. 

Next, we calculate the c.c.d.f. of ${\rm SIR}_{{\rm DASH},n,\ell}$. 
Similarly, there are two types of interferers, namely, i) interfering helpers storing 
version $\ell$ of DASH-based video $n$ (these helpers are further than the serving helper), 
and ii) interfering helpers not storing version $\ell$ of DASH-based video $n$ (these helpers could be closer to $u_0$ than the serving helper). By carefully handling these two types of interferers, we can calculate the c.c.d.f. of ${\rm SIR}_{{\rm DASH},n,\ell}$, using tools from stochastic geometry.

\begin{Lem}[c.c.d.f. of SIR]\label{Lem:distribution_SIR_dash}
The c.c.d.f. of ${\rm SIR}_{{\rm DASH},n,\ell}$ is given by $\Pr[{\rm SIR}_{{\rm DASH},n,\ell}\geq\tau]=f\left(\tau, T_{n,\ell}\right)$,
where $f(\tau,x)$ is given by \eqref{eqn:f}.
\end{Lem}

\begin{IEEEproof}
Lemma~\ref{Lem:distribution_SIR_dash} can be proved in a similar way to Lemma~\ref{Lem:distribution_SIR}. We omit the details due to page limitation.
\end{IEEEproof}

From Lemma~\ref{Lem:distribution_SIR_dash}, we can see that the impact of the physical layer parameters $\alpha$ and $\tau$ (captured by \textcolor{black}{$D_1(\tau) $ and $D_2(\tau)$}) and the impact of the caching distribution $\mathbf p$ on the c.c.d.f. of ${\rm SIR}_{{\rm DASH},n,\ell}$ are separated.
In addition, different from the random caching for SVC-based videos, $\Pr[{\rm SIR}_{{\rm DASH},n,\ell}\geq\tau]$ depends only on the probability that a helper stores version $\ell$ of DASH-based video $n$, i.e., $T_{n,\ell}$.

Based on Lemma~\ref{Lem:pmf_K_dash_n_x} and Lemma~\ref{Lem:distribution_SIR_dash}, we can obtain 
$q_{\rm DASH}(\mathbf p)$ as follows.

\begin{Thm}[Performance for DASH-based Videos]\label{Thm:STP_DASH}
The successful transmission probability for DASH-based videos is given by
\begin{align}
q_{\rm DASH}(\mathbf p) = \sum_{n\in\mathcal N}a_{n}\sum_{\ell\in\mathcal L} b_{n,\ell}\sum_{k\in\mathcal K_{{\rm DASH},n,\ell}}\Pr[K_{{\rm DASH},n,\ell}=k]\Pr[{\rm SIR}_{{\rm DASH},n,\ell}\geq\tau_k],\notag
\end{align}
where $\Pr[K_{{\rm DASH},n,\ell}=k]$ is given by Lemma~\ref{Lem:pmf_K_dash_n_x} and $\Pr[{\rm SIR}_{{\rm DASH},n,\ell}\geq\tau_k]$ is given by Lemma~\ref{Lem:distribution_SIR_dash}.
\end{Thm}

To obtain design insights into caching and multicasting for DASH-based videos, we analyze the successful transmission probability in the high user density region. Note that when $\textcolor{black}{C}\gg S_{{\rm DASH},L}$, by \eqref{eqn:x_n_l_larger_K_dash}, we know that $\frac{\textcolor{black}{C}-\sum_{n\in\mathcal N}\sum_{\ell\in\mathcal L}x_{n,\ell}S_{{\rm DASH},\ell}}{\textcolor{black}{C}}\approx 0$ for all $\mathbf x\in\mathcal X_{\rm DASH}$.
Approximating $\sum_{n\in\mathcal N}\sum_{\ell\in\mathcal L}x_{n,\ell}S_{{\rm DASH},\ell}$ with $\textcolor{black}{C}$, from Theorem~\ref{Thm:STP_DASH}, we have the following corollary.

\begin{Cor}[Performance for DASH-based Videos when $\lambda_u\to \infty$]\label{Cor:STP_DASH_asymp}
\begin{align}
 \lim_{\lambda_u\to \infty} q_{\rm DASH}(\mathbf p)=\sum_{n\in\mathcal N} \sum_{\ell\in\mathcal L} a_nb_{n,\ell}f(\tau_{\textcolor{black}{C}},T_{n,\ell})\triangleq q_{\rm DASH,\infty}(\mathbf T),\notag
\end{align}
where $f(\tau,x)$ is given by \eqref{eqn:f}.
\end{Cor}

From Corollary~\ref{Cor:STP_DASH_asymp}, we can see that the successful transmission probability of version $\ell$ of DASH-based video $n$ in the high user density region $f(\tau_C,T_{n,\ell})$ is a  concave increasing function of $T_{n,\ell}$. In addition, $q_{{\rm DASH},\infty}(\mathbf T)$ is a concave increasing function of $\mathbf T$, since $\frac{x}{D_2(\tau_C)+D_1(\tau_C)x}$ is a concave function of $x$ and a nonnegative weighted sum of concave functions is concave.
Fig.~\ref{fig:SVC_anal_vs_MC_vs_asymp}~(b) plots the successful transmission probability for DASH-based videos versus the user density $\lambda_u$. Similarly, Fig.~\ref{fig:SVC_anal_vs_MC_vs_asymp}~(b) verifies Theorem~\ref{Thm:STP_DASH} and Corollary~\ref{Cor:STP_DASH_asymp}. 


\subsection{Performance Optimization for DASH-based Videos}

In this part, we optimize the caching distribution $\mathbf p$ to maximize the successful transmission probability $q_{{\rm DASH},\infty}(\mathbf T)$ in the high user density region. \textcolor{black}{Similarly, we shall see that the obtained solution also achieves promising performance in the general user density region}.

\subsubsection{Problem Formulation}
We would like to maximize $q_{{\rm DASH},\infty}(\mathbf T)$ by carefully optimizing $\mathbf p$ under the constraints in \eqref{eqn:constraint_0_1} and \eqref{eqn:constraint_sum}. 

\begin{Prob}[Random Caching and Multicasting for DASH-based Videos]\label{Prob:DASH}
\begin{align}
q^*_{\rm DASH}\triangleq \max_{\mathbf p}  &  \quad  q_{{\rm DASH},\infty}(\mathbf T)\notag\\
s.t. \ & \quad \eqref{eqn:constraint_0_1}, \eqref{eqn:constraint_sum}.\notag
\end{align}
\end{Prob}

Similarly to Problem~\ref{Prob:SVC}, Problem~\ref{Prob:DASH} is a convex problem, but
cannot be solved with acceptable \textcolor{black}{computational} complexity for a practical network where $N$ and $\textcolor{black}{C}$ are usually very large.

\subsubsection{Near Optimal Solution}
We adopt a two-stage optimization method, similar to the one in Section~\ref{subsec:opt_SVC}, to obtain a low-complexity near optimal solution of Problem~\ref{Prob:DASH}.
In Stage I, we formulate and solve a relaxed problem of Problem~\ref{Prob:DASH} with $\mathbf T$ as the optimization variable instead of $\mathbf p$. Specifically, by~\eqref{eqn:x_n_l_K_dash}, \eqref{eqn:constraint_0_1}, \eqref{eqn:constraint_sum} and \eqref{eqn:svc_relation_T_p}, we have
\begin{align}
&0\leq T_{n,\ell} = \sum_{{\mathbf x}\in\mathcal X_{\rm DASH}:x_{n,\ell}=1}p_{\mathbf x}\leq \sum_{\mathbf x\in\mathcal X_{\rm DASH}}p_{\mathbf x}=1,\quad n\in\mathcal N, \ell\in\mathcal L,\label{eqn:dash_constraint_Tn_0_1}\\
&\sum_{n\in\mathcal N}\sum_{\ell\in\mathcal L}S_{{\rm DASH},\ell} T_{n,\ell}=\sum_{n\in\mathcal N}\sum_{\ell\in\mathcal L}S_{{\rm DASH},\ell} \sum_{\mathbf x\in\mathcal X_{\rm DASH}}p_{\mathbf x}x_{n,\ell}=\sum_{\mathbf x\in\mathcal X_{\rm DASH}}p_{\mathbf x}\sum_{n\in\mathcal N}\sum_{\ell\in\mathcal L}S_{{\rm DASH},\ell}x_{n,\ell}\leq \textcolor{black}{C}.\label{eqn:dash_constraint_sum_Tn_leq_K}
\end{align}
Therefore,
Problem~\ref{Prob:DASH} can be relaxed to \textcolor{black}{the following one}.

\begin{Prob}[Relaxed Problem of Problem~\ref{Prob:DASH}]\label{Prob:DASH_relaxed}
\begin{align}
q^*_{{\rm DASH},{\rm ub}}\triangleq \max_{\mathbf T}  &  \quad  q_{\rm DASH,\infty}(\mathbf T) \notag\\
s.t. \ &\quad \eqref{eqn:dash_constraint_Tn_0_1},\eqref{eqn:dash_constraint_sum_Tn_leq_K}.\notag
\end{align}
Let $\mathbf T_{\rm DASH}^*\triangleq (T_{{\rm DASH},n,\ell}^*)_{n\in\mathcal N, \ell\in\mathcal L}$ denote an optimal solution of Problem~\ref{Prob:DASH_relaxed}.
\end{Prob}

The optimal value of Problem~\ref{Prob:DASH_relaxed} can serve as an upper bound for that of Problem~\ref{Prob:DASH}, i.e., $q_{\rm DASH}^*\leq q^*_{{\rm DASH},ub}$. Note that the number of optimization variables in Problem~\ref{Prob:DASH_relaxed} is $NL$ which is much smaller than that in Problem~\ref{Prob:DASH}.
In addition, it can be easily seen that Problem~\ref{Prob:DASH_relaxed} is a convex problem and Slater's condition is satisfied, implying that strong duality holds. Using KKT conditions, we can solve Problem~\ref{Prob:DASH_relaxed}.

\begin{Lem}[Optimal Solution of Problem~\ref{Prob:DASH_relaxed}]\label{Lem:DASH_opt}
The optimal solution of Problem~\ref{Prob:DASH_relaxed} is given by
\begin{align}
T_{{\rm DASH},n,\ell}^*=\min\left\{\left[\frac{1}{D_1(\tau_C)}\sqrt{\frac{a_nb_{n,\ell}D_2(\tau_C)}{v^*S_{{\rm DASH},\ell}}}-\frac{D_2(\tau_C)}{D_1(\tau_C)}\right]^+,1\right\}, \quad n\in\mathcal N,\ l\in\mathcal L,
\end{align}
where 
$v^*$ satisfies $\sum_{n\in\mathcal L}\sum_{\ell\in\mathcal L}\min\left\{\left[\frac{1}{D_1(\tau_C)}\sqrt{\frac{a_nb_{n,\ell}D_2(\tau_C)}{v^*S_{{\rm DASH},\ell}}}-\frac{D_2(\tau_C)}{D_1(\tau_C)}\right]^+,1\right\}S_{{\rm DASH},\ell}=\textcolor{black}{C}$.
\end{Lem}

Similar to the optimal solution for independent single-quality files in Theorem $2$ of~\cite{Cui16TWC}, the optimal solution in Lemma~\ref{Lem:DASH_opt} for DASH-based videos has a reverse water-filling structure. Furthermore, when $L=1$, the optimal solution in Lemma~\ref{Lem:DASH_opt} reduces to the one in Theorem 2 of~\cite{Cui16TWC}. 
Based on Lemma~\ref{Lem:DASH_opt}, we have the following result.
\begin{Lem}[Optimality Property of Problem~\ref{Prob:DASH_relaxed}]\label{Lem:dash_property_opt_T}
For all $n_1,n_2\in\mathcal N$ and for all $\ell,j\in\mathcal L$, if $\frac{a_{n_1}b_{n_1,\ell}}{S_{{\rm DASH},\ell}}\geq \frac{a_{n_2}b_{n_2,j}}{S_{{\rm DASH},j}}$, then $T_{{\rm DASH},n_1,\ell}^*\geq T_{{\rm DASH},n_2,j}^*$.
\end{Lem}

Lemma~\ref{Lem:dash_property_opt_T} indicates that the caching probability of version $\ell$ of DASH-based video $n_1$ is no smaller than that of version $j$ of DASH-based video $n_2$, if $\frac{a_{n_1}b_{n_1,\ell}}{S_{{\rm DASH},\ell}}\geq \frac{a_{n_2}b_{n_2,j}}{S_{{\rm DASH},j}}$. This is because storing version $\ell$ of DASH-based video $n_1$ can satisfy more requests per unit storage than storing version $j$ of DASH-based video $n_2$.
Note that Lemma~\ref{Lem:dash_property_opt_T} implies that if $\frac{a_{n_1}b_{n_1,\ell}}{S_{{\rm DASH},\ell}}= \frac{a_{n_2}b_{n_2,j}}{S_{{\rm DASH},j}}$, then $T_{{\rm DASH},n_1,\ell}^*= T_{{\rm DASH},n_2,j}^*$, as $T_{{\rm DASH},n_1,\ell}^*= T_{{\rm DASH},n_2,j}^*$ is equivalent to $T_{{\rm DASH},n_1,\ell}^*\geq T_{{\rm DASH},n_2,j}^*$ and $T_{{\rm DASH},n_1,\ell}^*\leq  T_{{\rm DASH},n_2,j}^*$.
In addition, note that in Lemma~\ref{Lem:dash_property_opt_T}, we allow $n_1=n_2$ or $\ell=j$. In particular, when $n_1=n_2$, Lemma~\ref{Lem:dash_property_opt_T} implies that if $\frac{b_{n,\ell}}{S_{{\rm DASH},\ell}}\geq \frac{b_{n,j}}{S_{{\rm DASH},j}}$ (note that $\frac{b_{n,\ell}}{S_{{\rm DASH},\ell}}$ can be interpreted as the caching gain \textcolor{black}{of} description $\ell$ \textcolor{black}{of} DASH-based video $n$), then $T_{{\rm DASH},n,\ell}^*\geq T_{{\rm DASH},n,j}^*$; 
when $\ell=j$, Lemma~\ref{Lem:dash_property_opt_T} implies that if $a_{n_1}b_{n_1,\ell}\geq a_{n_2}b_{n_2,\ell}$, then $T_{{\rm DASH},n_1,\ell}^*\geq T_{{\rm DASH},n_2,\ell}^*$. 

\begin{algorithm} \caption{Near Optimal Solution of Problem~\ref{Prob:DASH} ($g={\rm DASH}$)}
\small{\begin{algorithmic}[1]
\STATE Initialize $\mathcal A = [0,1]\times[0,\textcolor{black}{C}]$, $\mathcal I=[0,1]$, $\mathcal O_{n,\ell}=\emptyset$ for all $n\in\mathcal N$ and $\ell\in\mathcal L$,  $\mathcal P= \{0,1\}$, and $\mathbf R=\mathbf T_{g}^*$.
\STATE Sort  $\mathbf R$ in decreasing order. Let $R_{(1),<1>}\geq \cdots \geq R_{(NL),<NL>}$ be the resulting sequence. 
\FOR{$k=1:NL$}
\WHILE{$R_{(k),<k>}>0$}
\STATE Set $y_b=\min\{y:(x,y)\in\mathcal A,x\in \mathcal I\}$.
\IF{$y_b\leq K-S_{{g},<k>}$}
\STATE Set $x_l=\min\{x\in \mathcal I: (x,y_b)\in\mathcal A\}$ and $x_r=\max\{x\in \mathcal I: a(x_l,y_b)+(1-a)(x,y_b)\in \mathcal A\ \text{for all}\ a\in[0,1]\}$.
\IF {$R_{(k),<k>}\leq x_r-x_l$}
\STATE Set $\mathcal O_{(k),<k>}=\mathcal O_{(k),<k>}\cup [x_l,x_l+R_{(k),<k>}]$, $\mathcal A = \mathcal A\setminus ([x_l,x_l+R_{(k),<k>}]\times[y_b,y_b+S_{{g},<k>}])$, $\mathcal P= \mathcal P\cup \{x_l+R_{(k),<k>}\}$ and $R_{(k),<k>}=0$.
\ELSE
\STATE Set $\mathcal O_{(k),<k>}=\mathcal O_{(k),<k>}\cup [x_l,x_r]$, $\mathcal A = \mathcal A\setminus ([x_l,x_r]\times[y_b,y_b+S_{{g},<k>}])$, $\mathcal P= \mathcal P\cup \{x_r\}$, and $R_{(k),<k>}=R_{(k),<k>}-(x_r-x_l)$.
\ENDIF
\ELSE
\STATE Set $R_{(k),<k>}=0$.
\ENDIF
\ENDWHILE
\ENDFOR
\STATE Same as Step 20 as Algorithm~\ref{alg:SVC_opt}. 
\FOR{$i=1:|\mathcal P|-1$}
\STATE Same as Step 22 as Algorithm~\ref{alg:SVC_opt}. 
\IF{$j\in\mathcal L$}
\STATE Set $m=\mathop{\arg\max}\limits_{n\in\mathcal N:[x_{(i)},x_{(i+1)}]\cap \mathcal O_{n,j}=\emptyset}a_{n} b_{n,j}(f(\tau_C,T_{{g},n,j}^{\dagger}+x_{(i+1)}-x_{(i)})-f(\tau_C,T_{{g},n,j}^{\dagger}))$, $\mathcal O_{m,j}=\mathcal O_{m,j}\cup [x_{(i)},x_{(i+1)}]$, and $T_{{g},m,j}^{\dagger}=T_{{g},m,j}^{\dagger}+x_{(i+1)}-x_{(i)}$.
\ENDIF
\ENDFOR
\STATE Same as Steps 27-34 as Algorithm~\ref{alg:SVC_opt}.
\end{algorithmic}}\normalsize\label{alg:DASH_opt}
\end{algorithm}

In Stage II, based on the optimal solution $\mathbf T_{\rm DASH}^*$ of Problem~\ref{Prob:DASH_relaxed}, we construct a joint distribution which serves as a near optimal solution of Problem~\ref{Prob:DASH}, as summarized in Algorithm~\ref{alg:DASH_opt}. An illustration example is shown in Fig.~\ref{fig:form_combination}~(b). \textcolor{black}{Note that when $L=1$, Algorithm~\ref{alg:DASH_opt} reduces to the probabilistic caching method in~\cite{ICC15Giovanidis}. Similarly, we can show $\parallel\mathbf T_{\rm DASH}^*-\mathbf T_{\rm DASH}^\dagger\parallel_1<2$ and  $\frac{\parallel \mathbf T_{\rm DASH}^*-\mathbf T_{\rm DASH}^\dagger\parallel_1}{\parallel \mathbf T_{\rm DASH}^* \parallel_1}< \frac{2S_{{\rm DASH},L}}{C}$ (which indicates that the relative difference between $\mathbf T_{\rm DASH}^*$ and $\mathbf T_{\rm DASH}^\dagger$ is negligible when $\textcolor{black}{C}\gg S_{{\rm DASH},L}$). In addition, we can show that $\mathbf p_{\rm DASH}^\dagger$ is an optimal solution of Problem~\ref{Prob:DASH} when $L\geq 2$ and $\sum_{n\in\mathcal N}\sum_{\ell\in\mathcal L}R_{n,\ell}=0$ and when $L=1$}. 

In Fig.~\ref{fig:opt_ub_lb}~(b) and Table~\ref{tab:secondtable}, we use a numerical example to compare the optimal solution of Problem~\ref{Prob:DASH} obtained by CVX and the proposed near optimal solution obtained by the two-stage optimization method for DASH-based videos in both successful transmission probability and \textcolor{black}{computational} complexity.
Fig.~\ref{fig:opt_ub_lb}~(b) and and Table~\ref{tab:secondtable} demonstrate the applicability and effectiveness of the near optimal solution.

\section{Numerical Results}


\begin{table}[t]
\caption{\textcolor{black}{SVC video encoding parameters.}}\label{tab:para_SVC}
\begin{center}
\vspace{-6mm}
\begin{scriptsize}
\textcolor{black}{
\begin{tabular}{ ll }
\hline
Parameter name & Parameter value \\
\hline
Video codec & H.264/Scalable Video Coding\\
Video sequence & City (YUV CIF $352\times288$ pixels @ $30$ fps)\\
FramesToBeEncoded & $300$\\
No. of layers & $3$ (1 base layer, 2 enhancement layers) \\
GoPsize &  $4$\\
IntraPeriod & $16$ frames\\
FrameRateOut & $15$, $15$, $30$\\
SourceWidth $\times$ SourceHeight & $176\times144$, $352\times288$,  $352\times288$\\
QP & $32$, $36$,  $36$\\
\hline
\end{tabular}}
\end{scriptsize}
\vspace{-8mm}
\end{center}
\end{table}

\begin{table}[t]
\caption{\textcolor{black}{DASH video encoding parameters.}}\label{tab:para_AVC}
\begin{center}
\vspace{-6mm}
\begin{scriptsize}
\textcolor{black}{
\begin{tabular}{ ll }
\hline
Parameter name & Parameter value \\
\hline
Video codec & H.264/Advance video coding\\
Video sequence & City (YUV CIF $352\times288$ pixels @ $30$ fps)\\
FramesToBeEncoded & $150$, $150$, $300$\\
OutputWidth $\times$ OutputHeight &  $176 \times144$, $352 \times288$, $352 \times288$\\
FrameSkip & $1$, $1$, $0$\\
IntraPeriod & $8$, $8$,  $16$ frames\\
\hline
\end{tabular}}
\end{scriptsize}
\vspace{-8mm}
\end{center}
\end{table}

\begin{table}[t]
\caption{\textcolor{black}{Outputs of DASH and SVC video encoding.}}\label{tab:Output_DASH_SVC}
\begin{center}
\vspace{-6mm}
\begin{scriptsize}
\textcolor{black}{
\begin{tabular}{|c|c|c|c|c|c|c|}
\hline
Version $\ell$ & Resolution@FPS & Bitrate of SVC (Kbps) & Bitrate of DASH (Kbps)\\
\hline
$1$ & $176\times144$@$15$ & $67.8344$ & $56.60$ \\
\hline
$2$ & $352\times288$@$15$ & $178.5712$  & $143.34$ \\
\hline
$3$ & $352\times288$@$30$ & $288.4632$ & $227.64$  \\
\hline
\end{tabular}}
\end{scriptsize}
\vspace{-8mm}
\end{center}
\end{table}


In this section, we compare the proposed near optimal solutions for SVC-based videos and DASH-based videos with \textcolor{black}{four baseline schemes for SVC-based videos and four baseline schemes for DASH-based videos, respectively}. 
Most popular-SVC-ver. $\ell$ ($\ell\in \{1,L\}$) refers to the caching design where each helper stores version $\ell$ of each of the $\lfloor\frac{C}{S_{{\rm SVC},\ell}}\rfloor$ most popular SVC-based videos~\cite{Zhang17GLOBECOM}.
Uniform dist.-SVC-ver. $\ell$ ($\ell\in \{1,L\}$) refers to the caching design where each helper stores version $\ell$ of an SVC-based video in $\mathcal N$ chosen uniformly at random (i.e., with probability $\frac{C}{NS_{{\rm SVC},\ell}}$).
\textcolor{black}{Most popular-DASH-$L$ ver.s refers to the caching design where each helper stores all $L$ versions of each of the $\lfloor\frac{C}{\sum_{\ell\in\mathcal L}S_{{\rm DASH},\ell}}\rfloor$ most popular DASH-based videos. Uniform dist.-DASH-$L$ ver.s refers to the caching design where each helper stores all $L$ versions of a DASH-based video in $\mathcal N$ chosen uniformly at random (i.e., with probability $\frac{C}{N\sum_{\ell\in\mathcal L}S_{{\rm DASH},\ell}}$).}
Most popular-DASH-ver. $1$ refers to the caching design where each helper stores version $1$ of each of the $\lfloor\frac{C}{S_{{\rm DASH},1}}\rfloor$ most popular DASH-based videos.
Uniform dist.-DASH-ver. $1$ refers to the caching design where each helper stores version $1$ of a DASH-based video in $\mathcal N$ chosen uniformly at random (i.e., with probability $\frac{C}{NS_{{\rm DASH},1}}$).
\textcolor{black}{Note that most popular-DASH-$L$ ver.s, uniform dist.-DASH-$L$ ver.s, most popular-DASH-ver. $1$ and uniform dist.-DASH ver.~$1$ can be viewed as counterparts of most popular-SVC-ver. $L$, uniform dist.-SVC ver.~$L$, most popular-SVC-ver. $1$ and uniform dist.-SVC ver.~$1$, respectively}.

In the simulation, \textcolor{black}{we consider a real video source ``City''~\cite{VideoSource} and adopt JSVM and JM as the SVC encoder and DASH encoder, respectively. Tables~\ref{tab:para_SVC} and~\ref{tab:para_AVC} summarize the encoding parameters for the SVC-based video and the DASH-based video, respectively. The encoding results are given in Table~\ref{tab:Output_DASH_SVC}. 
In particular, when encoding the SVC-based video with JSVM, we consider the spatial scalable coding, which supports multiple spatial resolutions. For a fair comparison, we then adjust the quantization parameter in JM to ensure that the PSNR of the DASH-based video is approximately the same as that of the SVC-based video}. 

For ease of exposition, we assume 
$\mathbf a$ and $\mathbf b_n$, $n\in\mathcal N$ follow Zipf distributions with Zipf exponents $\gamma_1$ and $\gamma_2$, respectively, i.e., $a_{n}=\frac{n^{-\gamma_1}}{\sum_{n\in\mathcal N}n^{-\gamma_1}}$ (with $n$  being also the rank index) and $b_{n,\ell}=\frac{\ell^{-\gamma_2}}{\sum_{\ell\in\mathcal L}\ell^{-\gamma_2}}$ (with $\ell$  being also the rank index). 
Unless otherwise stated, 
we choose $N=100$, $L=3$, $\textcolor{black}{C}=60$, \textcolor{black}{$\gamma_1=0.8$, $\gamma_2=1.4$}, \textcolor{black}{$R_{{\rm SVC},1}=67.8344$ Kbps, $R_{{\rm SVC},2}=178.5712$ Kbps, $R_{{\rm SVC},3}=288.4632$ Kbps, $R_{{\rm DASH},1}=56.60$ Kbps, $R_{{\rm DASH},2}=143.34$ Kbps, $R_{{\rm DASH},3}=227.64$ Kbps}, 
\textcolor{black}{$S_{{\rm SVC},1}=1.2$, $S_{{\rm SVC},2}=3.15$, $S_{{\rm SVC},3}=5.1$, $S_{{\rm DASH},1}=1$, $S_{{\rm DASH},2}=2.53$, $S_{{\rm DASH},3}=4.02$}, \textcolor{black}{$W=20$} MHz and $\alpha=4$. 

\begin{figure}[t]
\begin{center}
\subfigure[\small{SVC-based videos.}]
{\resizebox{5.2cm}{!}{\includegraphics{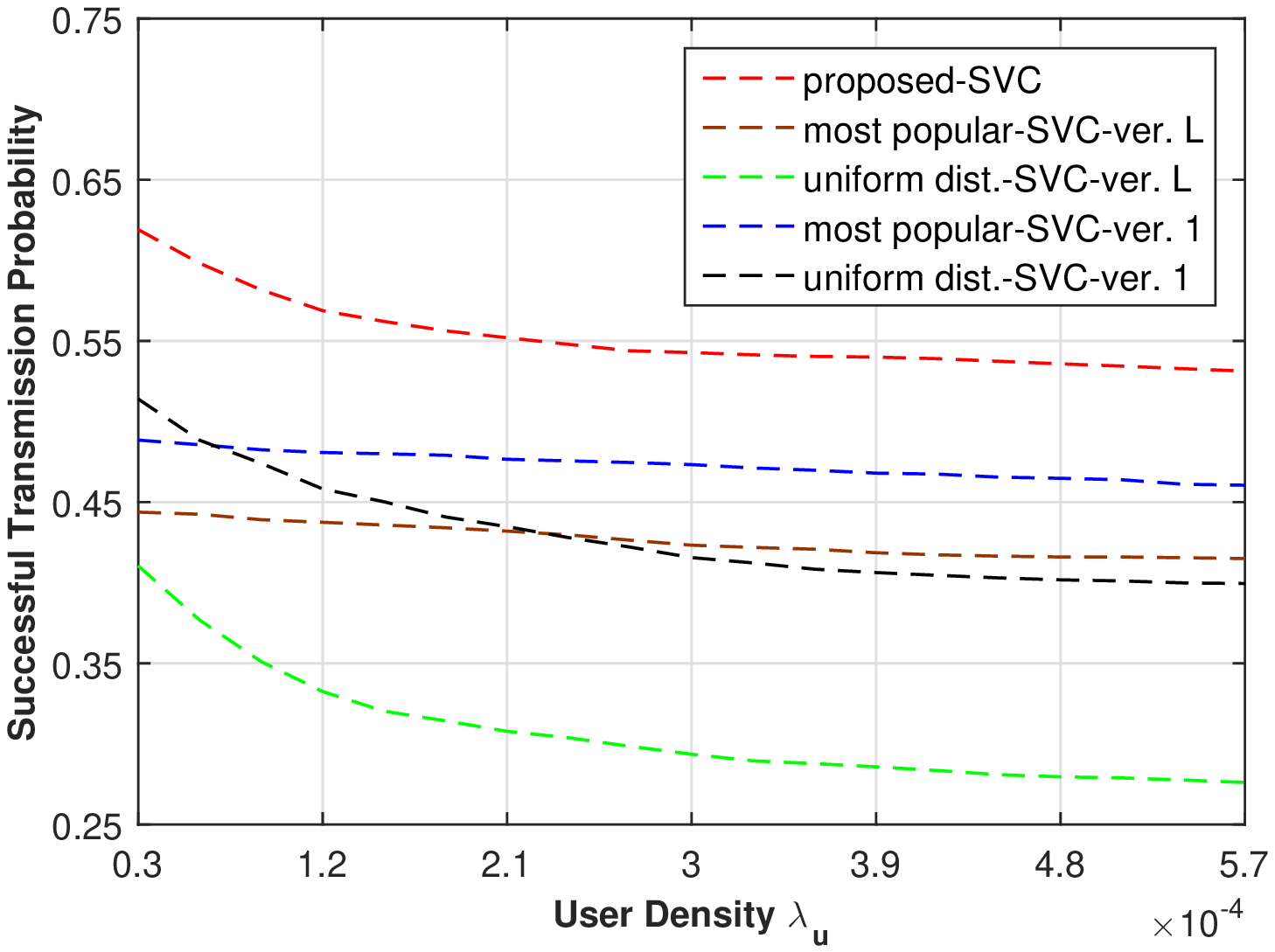}}}\hspace{10mm}
\subfigure[\small{DASH-based videos.}]
{\resizebox{5.2cm}{!}{\includegraphics{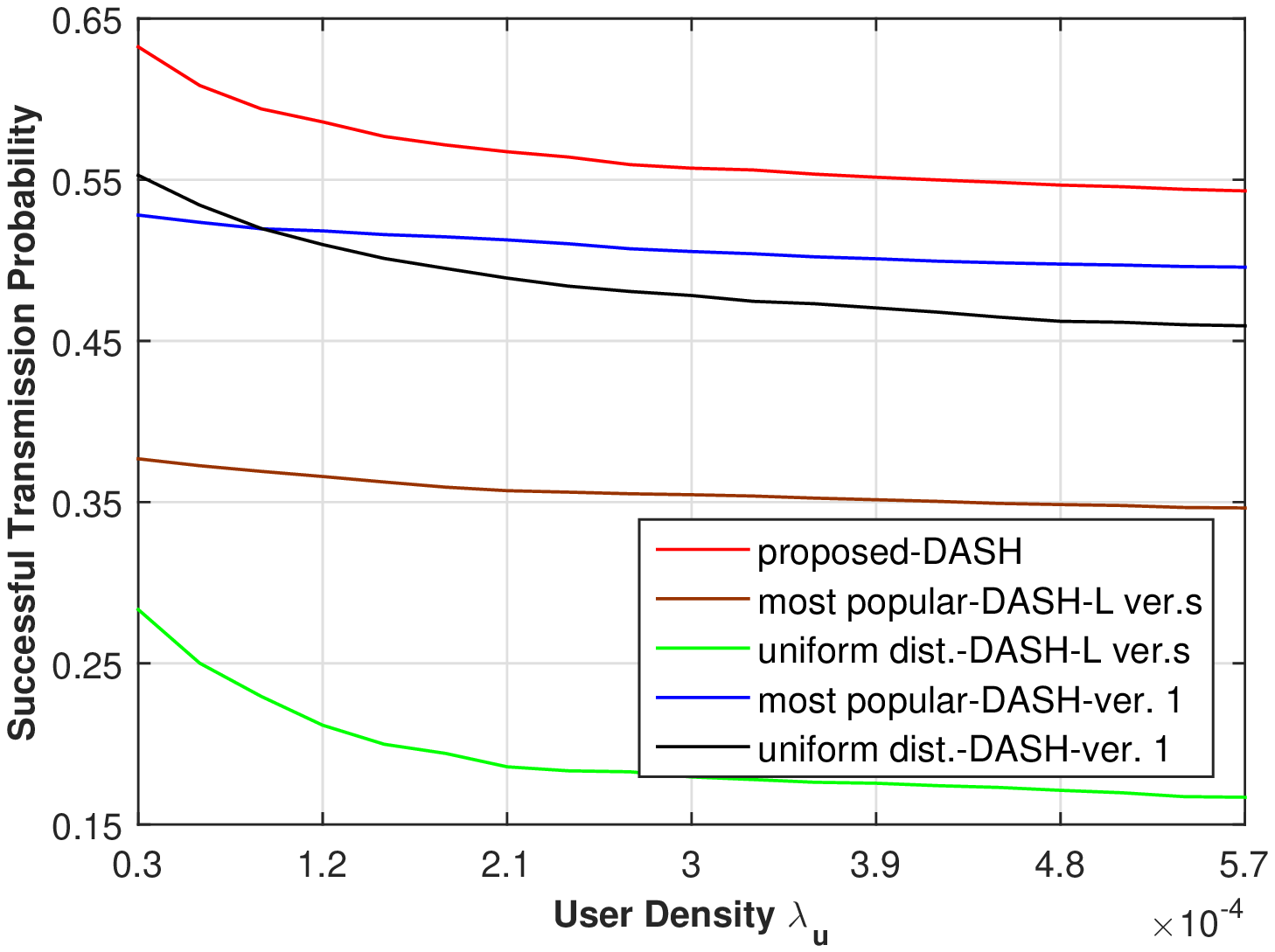}}}
\end{center}
\vspace{-2mm}
\caption{\textcolor{black}{\small Successful transmission probability versus user density $\lambda_u$. $\lambda_b=3\times 10^{-6}$.}}
\vspace{-4mm}
\label{fig:STP_vs_lambda_u}
\end{figure}

\textcolor{black}{Fig.~\ref{fig:STP_vs_lambda_u} plots the successful transmission probability of each  scheme versus user density $\lambda_u$ in the general user density region.  From Fig.~\ref{fig:STP_vs_lambda_u}, we can see that the proposed solutions outperform the corresponding baseline schemes under the considered user densities.  This indicates that the proposed solutions, although obtained for the high user density region, can effectively exploit the storage resource over the entire range of user density of interest, as they can capture the key impacts of the other system parameters on the successful transmission probability and the impact of the user density on the successful transmission probability is not a dominant one.}

\begin{figure}[t]
\begin{center}
\subfigure[\small{\textcolor{black}{Proposed solutions for SVC-based and DASH-based videos}.}]
{\resizebox{5.2cm}{!}{\includegraphics{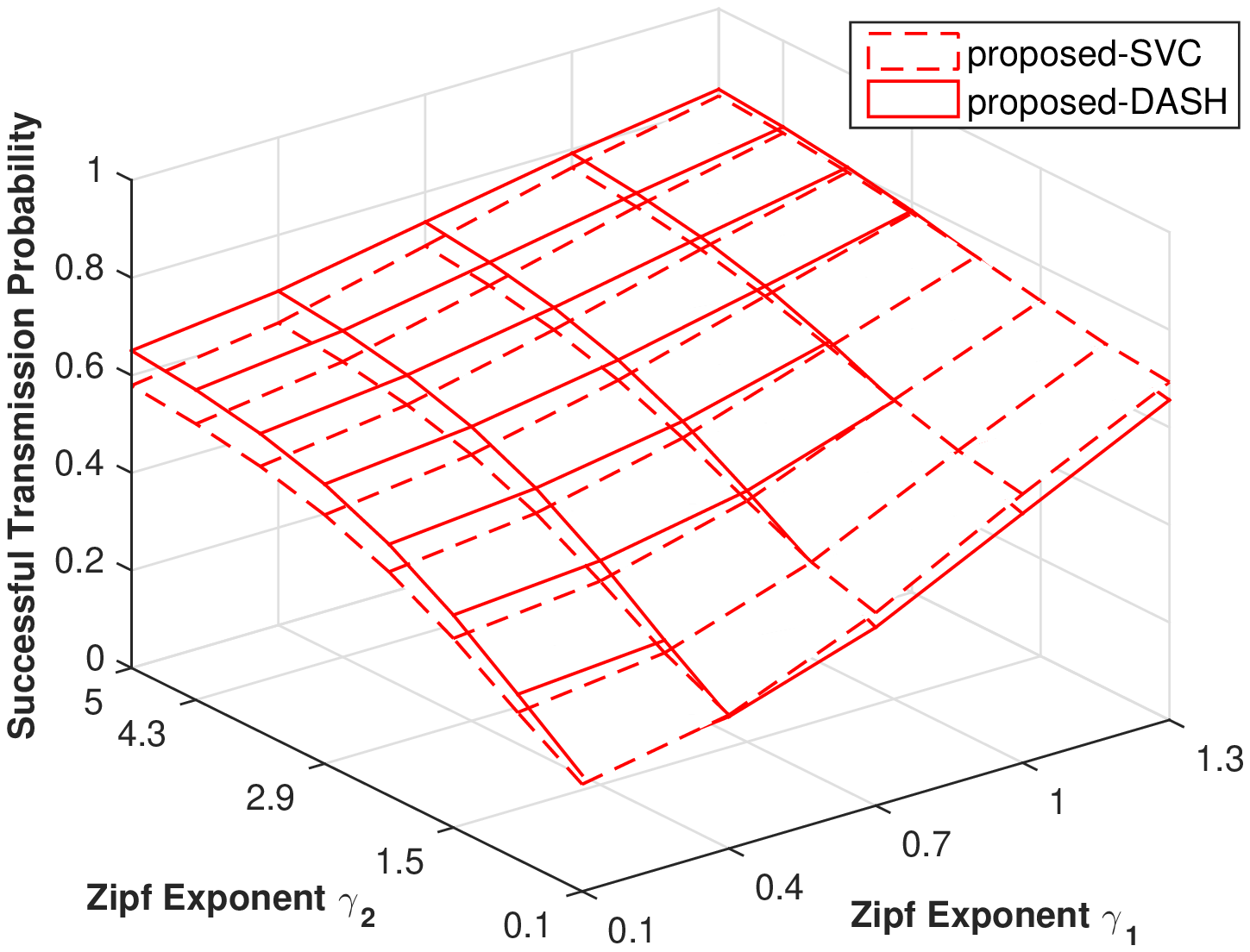}}}\quad
\subfigure[\small{\textcolor{black}{Baseline schemes storing all $L$ versions of an SVC-based video}.}]
{\resizebox{5.2cm}{!}{\includegraphics{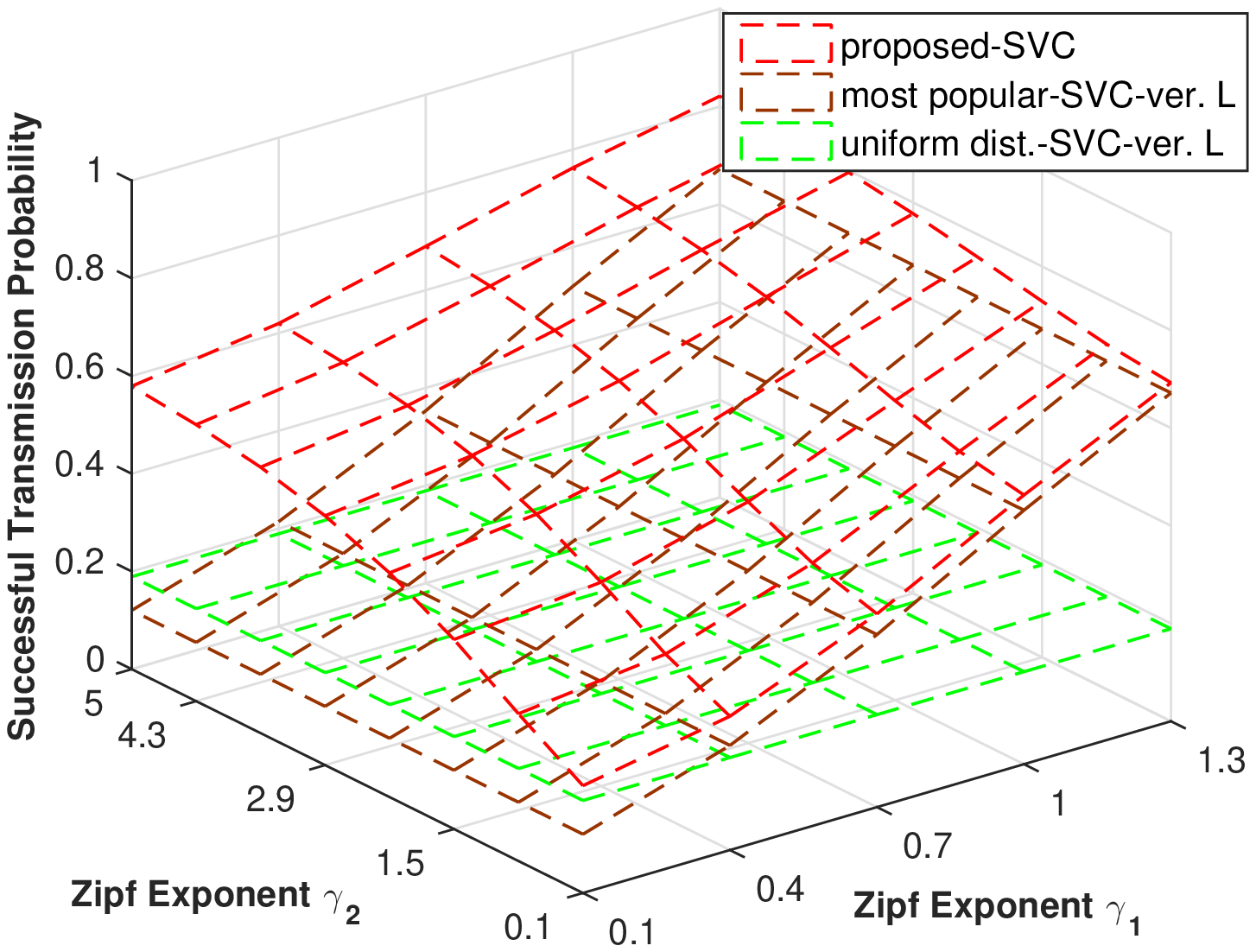}}}\quad
\subfigure[\small{\textcolor{black}{Baseline schemes storing all $L$ versions of a DASH-based video.}}]
{\resizebox{5.2cm}{!}{\includegraphics{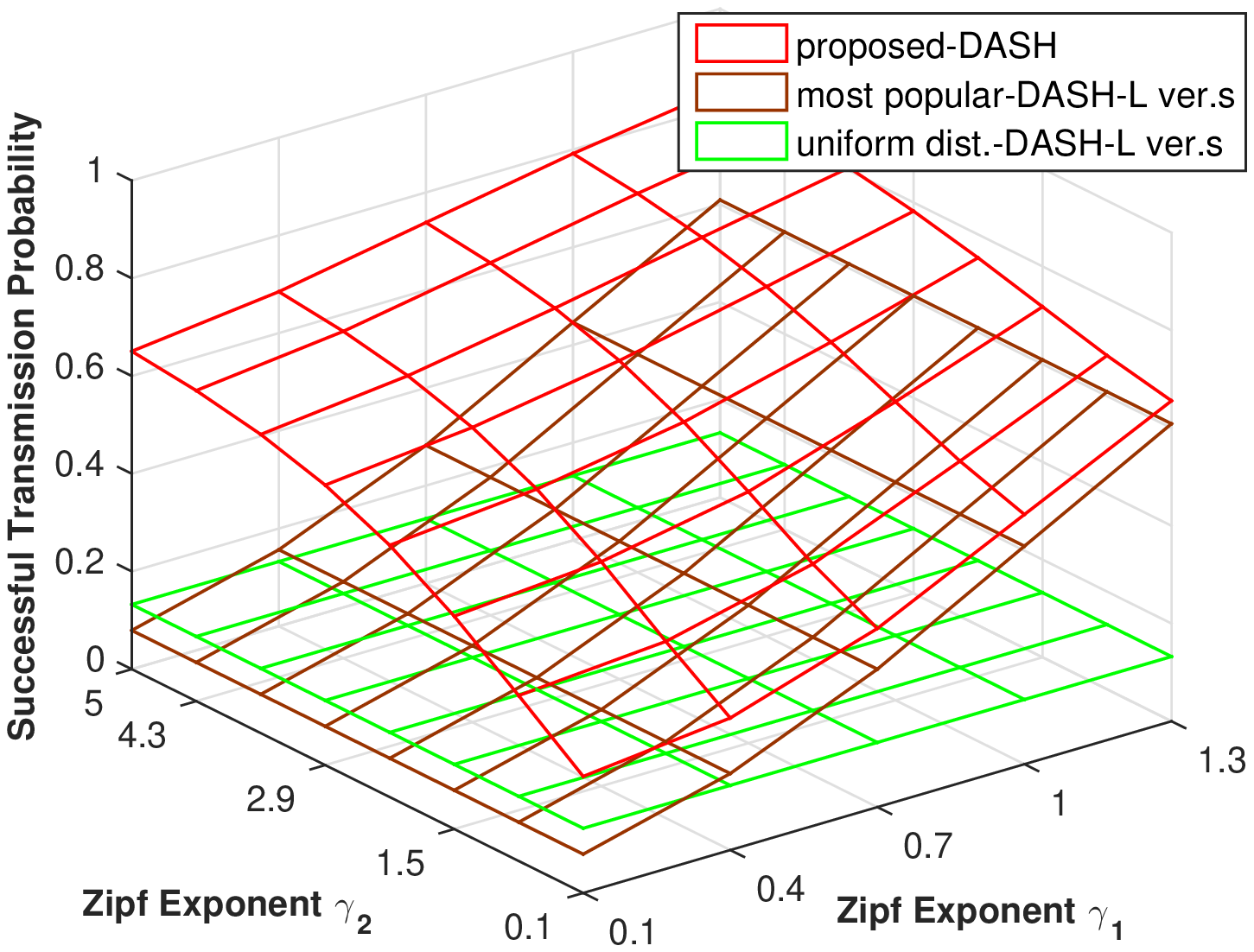}}}
\subfigure[\small{\textcolor{black}{Baseline schemes storing only version $1$ of an SVC-based video}}.]
{\resizebox{5.3cm}{!}{\includegraphics{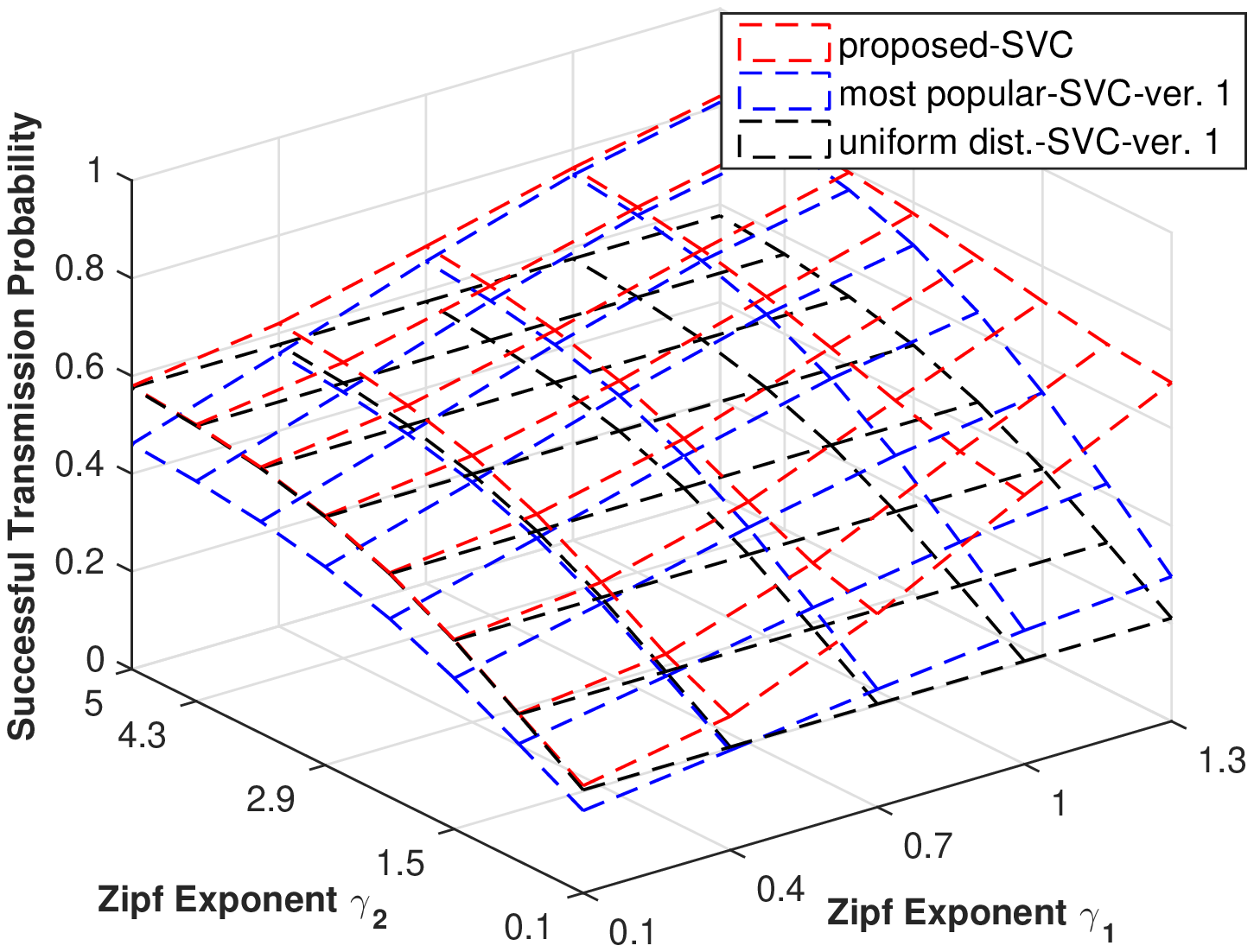}}}\quad
\subfigure[\small{\textcolor{black}{Baseline schemes storing only version $1$ of a DASH-based video}.}]
{\resizebox{5.3cm}{!}{\includegraphics{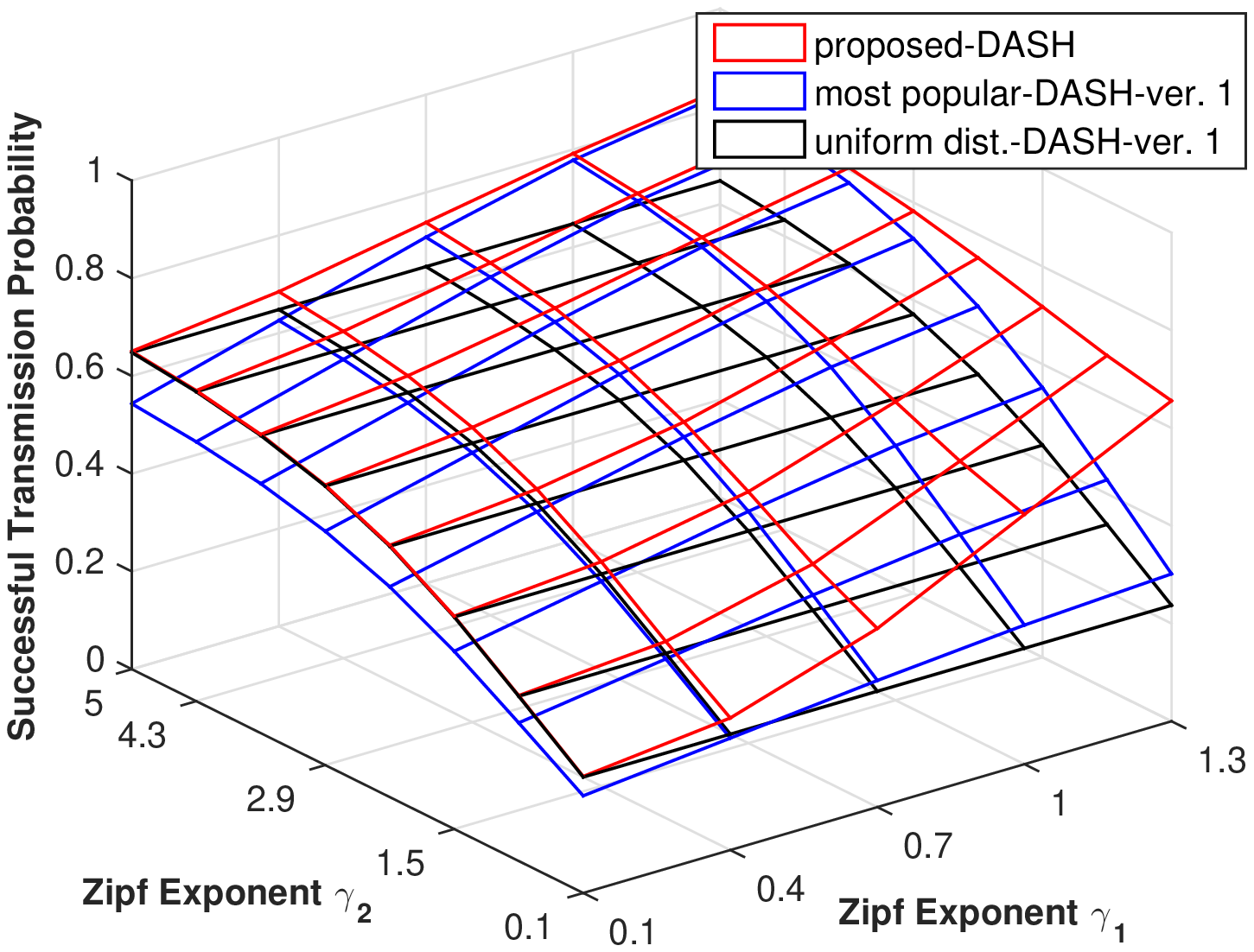}}}
\end{center}
\vspace{-2mm}
\caption{\small{Successful transmission probability versus Zipf exponents $\gamma_1$ and $\gamma_2$, and layered-encoding overhead $h$. }}
\vspace{-4mm}
\label{fig:performance_comparison}
\end{figure}

Fig.~\ref{fig:performance_comparison} (a) plots the successful transmission probability of each proposed solution versus \textcolor{black}{Zipf exponents $\gamma_1$ and $\gamma_2$}. 
We can observe that the proposed solutions for SVC-based videos and DASH-based videos \textcolor{black}{have different operating regions}. 
Specifically, 
the proposed solution for SVC-based videos outperforms that for DASH-based videos when $\gamma_1$ is large and $\gamma_2$ is small. 
\textcolor{black}{This is because} when $\gamma_1$ is large and $\gamma_2$ is small, the proposed solution for SVC-based videos stores more layers of the popular videos,  the proposed solution for DASH-based videos stores more descriptions of the popular videos, \textcolor{black}{and SVC provides more quality levels for given number of bits than the encoding in DASH, owing to its layered structure. In contrast, the proposed solution for SVC-based videos underperforms that for DASH-based videos when $\gamma_1$ is small and $\gamma_2$ is large. This is because when $\gamma_1$ is small and $\gamma_2$ is large, the proposed solution for SVC-based videos stores fewer layers of an SVC-based video, the proposed solution for DASH-based videos stores fewer descriptions of a DASH-based video, and SVC uses more bits to achieve the same quality level than the encoding in DASH, due to the layered encoding overhead}. 

Fig.~\ref{fig:performance_comparison} (b)-(e) plot the successful transmission probability of each scheme versus Zipf exponents $\gamma_1$ and $\gamma_2$. 
We can observe that the proposed solutions for SVC-based and DASH-based videos outperform the corresponding baseline schemes. 
This indicates that the proposed designs can effectively adapt to the popularity of multi-quality videos and wisely exploit the storage resource.
In addition, Fig.~\ref{fig:performance_comparison}~(b) shows 
that the performance of the proposed solution for SVC-based videos coincides with that of most popular-SVC-ver. $L$ when $\gamma_1$ is large and $\gamma_2$ is small, 
and coincides with that of uniform dist.-SVC-ver. $L$ when $\gamma_1$ and $\gamma_2$ are small, verifying Property (i) of Lemma~\ref{Lem:svc_property_opt_T}. 
\textcolor{black}{Fig.~\ref{fig:performance_comparison}~(c) shows that the performance gap between the proposed solution for DASH-based videos and most popular-DASH-$L$ ver.s decreases with $\gamma_1$ and increases with $\gamma_2$, and the performance gap between the proposed solution for DASH-based videos and uniform dist.-DASH-$L$ ver.s increases with $\gamma_1$ and $\gamma_2$}.
Fig.~\ref{fig:performance_comparison}~(d) (Fig.~\ref{fig:performance_comparison}~(e)) shows that 
the performance of the proposed solution for SVC-based (DASH-based) videos coincides with that of most popular-SVC-ver. $1$ (most popular-DASH-ver. $1$) when $\gamma_1$ and $\gamma_2$ are large, 
and coincides with that of uniform dist.-SVC-ver. $1$ (uniform dist.-DASH-ver. $1$) when $\gamma_1$ is small and $\gamma_2$ is large.
The reasons are as follows. For both types of videos, when $\gamma_1$ is large (small), storing \textcolor{black}{more} popular videos (more videos) can satisfy more requests. When $\gamma_2$ is large (small), storing fewer (more) layers of an SVC-based video can satisfy more requests. When $\gamma_2$ is large, storing fewer descriptions of a DASH-based video can satisfy more requests. 

\section{Conclusion}
In this paper, we considered the analysis and optimization of caching and multicasting for SVC-based videos and DASH-based videos respectively, in a large-scale wireless network.
First, we proposed a random caching and multicasting scheme for each type of videos. 
Then, for each type of videos, we derived tractable expressions for the successful transmission probability in the general and high user density regions respectively, utilizing tools from stochastic geometry.  
Next, for each type of videos, we considered the maximization of the successful transmission probability in the high user density region.
We proposed a two-stage optimization method to obtain a low-complexity near optimal solution, whose performance gap with \textcolor{black}{respect to an} optimal solution can be easily evaluated. Both analysis and optimization results reveal important design insights. 
Finally, \textcolor{black}{using numerical results based on real video sequences}, we showed that the proposed solution for each type of videos achieves a significant performance gain over baseline schemes \textcolor{black}{in the general and high user density regions}, \textcolor{black}{and demonstrated the respective operating regions of the proposed solutions for SVC-based and DASH-based videos}.


\section*{Appendix A: Proof of Lemma~\ref{Lem:pmf_K_svc_n_x}}

When $B_{{\rm SVC},n,\ell,0}$ stores cache content $\mathbf x$, we have $K_{{\rm SVC},n,\ell} = \sum_{m\in\mathcal N}S_{v_m}$ where $\mathbf v \preceq \mathbf x\mathbf z$ and $v_n\geq \ell$. Denote $\mathcal N_{\mathbf x}\triangleq \{n\in\mathcal N:\sum_{j\in\mathcal L}x_{n,j}=1\}$. We have 
\begin{align}
&\Pr[K_{{\rm SVC},n,\ell}=k| B_{{\rm SVC},n,\ell,0}\ \text{stores}\ \mathbf x] = \sum_{\mathbf k\in\mathcal {SQ}_{\mathbf x,n,\ell}(k)}\prod_{m\in\mathcal N_{\mathbf x}}\Pr[v_m=k_m| B_{{\rm SVC},n,\ell,0}\ \text{stores}\ \mathbf x].\label{eqn:pmf_K_n_l_app}
\end{align}
Now, we calculate $\Pr[v_m=k_m| B_{{\rm SVC},n,\ell,0}\ \text{stores}\ \mathbf x]$.
Let random variable $Y_{\mathbf x, n,\ell, m,j}\in\{0,1\}$ denote whether video $m$ with quality $j$ is requested by users associated with $B_{{\rm SVC},n,\ell,0}$ when $B_{{\rm SVC},n,\ell,0}$ stores cache content $\mathbf x$. We have 
\begin{align}
&\Pr[v_m=i| B_{{\rm SVC},n,\ell,0}\ \text{stores}\ \mathbf x] \notag\\
&=\begin{cases}
\prod_{j=1}^{u_m(\mathbf x)}\Pr[Y_{\mathbf x, n,\ell, m,j}=0],  &\quad m\in\mathcal N_{\mathbf x}\setminus\{n\}, i=0\\
\Pr[Y_{\mathbf x, n,\ell, m,i}=1]\prod_{j=i+1}^{u_m(\mathbf x)}\Pr[Y_{\mathbf x, n,\ell, m,j}=0], &\quad m\in\mathcal N_{\mathbf x}\setminus\{n\}, i=\{1,\cdots,u_m(\mathbf x)\}\\
0, & \quad m=n, i\in\{0,\cdots,\ell-1\}\\
\prod_{j=\ell+1}^{u_n(\mathbf x)}\Pr[Y_{\mathbf x, n,\ell, n,j}=0], &\quad m=n, i=\ell\\
\Pr[Y_{\mathbf x, n,\ell, n,i}=1]\prod_{j=i+1}^{u_n(\mathbf x)}\Pr[Y_{\mathbf x, n,\ell, n,j}=0], &\quad m=n, i=\{\ell+1,\cdots,u_n(\mathbf x)\}
\end{cases}.\label{eqn:pmf_y_m}
\end{align}
The p.m.f. of $Y_{\mathbf x, n,\ell, m,j}$ depends on the p.d.f. of the size of the Voronoi cell of $B_{{\rm SVC},n,\ell,0}$ w.r.t. video $m$ with quality $j$ when  $B_{{\rm SVC},n,\ell,0}$ stores cache content $\mathbf x$, which is unknown~\cite{SGcellsize13}. We approximate this p.d.f. based on the p.d.f. of the size of the Voronoi cell to which a randomly chosen user belongs~\cite{SGcellsize13}. Based on Lemma~$3$ of~\cite{SGcellsize13}, we have
\begin{align}
\Pr[Y_{\mathbf x, n,\ell, m,j}=0]=\left(1+\frac{a_{m}b_{m,j}\lambda_u}{3.5(\sum_{i=j}^LT_{m,i})\lambda_b}\right)^{-4.5}.\label{eqn:pmf_Y_v}
\end{align}
Based on \eqref{eqn:pmf_K_n_l_app}, \eqref{eqn:pmf_y_m} and \eqref{eqn:pmf_Y_v}, we can obtain $\Pr[K_{{\rm SVC},n,\ell}=k| B_{{\rm SVC},n,\ell,0}\ \text{stores}\ \mathbf x]$. Note that the probability that $B_{{\rm SVC},n,\ell,0}$ stores cache content $\mathbf x$ is
$\sum_{\mathbf x\in\mathcal X_{\rm SVC}:u_{n}(\mathbf x)\geq \ell}\frac{p_{\mathbf x}}{\sum_{j=\ell}^L{T_{n,j}}}$. Thus, by the law of total probability, we have $\Pr[K_{{\rm SVC},n,\ell}=k]=\sum_{\mathbf x\in\mathcal X_{\rm SVC}:u_{n}(\mathbf x)\geq \ell}\frac{p_{\mathbf x}}{\sum_{j=\ell}^L{T_{n,j}}}\Pr[K_{{\rm SVC},n,\ell}=k| B_{{\rm SVC},n,\ell,0}\ \text{stores}\ \mathbf x]$. Therefore, we complete the proof.

\section*{Appendix B: Proof of Lemma~\ref{Lem:distribution_SIR}}

First, we rewrite the ${\rm SIR}_{{\rm SVC},n,\ell}$ in \eqref{eqn:STP_SVC_def} as ${\rm SIR}_{{\rm SVC},n,\ell}=\frac{d_{0}^{-\alpha}|h_{0}|^2}{I_{n,\ell}+\overline{I}_{n,\ell}}$, where $\Phi_{h,n,\ell}$ denotes the point process generated by helpers storing at least the first $\ell$ layers of video $n$, $\overline{\Phi}_{h,n,\ell}\triangleq \Phi_h\setminus\Phi_{h,n,\ell}$, $I_{n,\ell}\triangleq \sum_{i\in\Phi_{h,n,\ell}\setminus \{B_{{\rm SVC},n,\ell,0}\}}d_i^{-\alpha}|h_i|^2$, and $\overline{I}_{n,\ell}\triangleq \sum_{i\in\overline{\Phi}_{h,n,\ell}}d_i^{-\alpha}|h_i|^2$. Due to the  independent thinning, point processes $\Phi_{h,n,\ell}$ and $\overline{\Phi}_{h,n,\ell}$ are two independent PPPs with density $\lambda_h\sum_{j=\ell}^L T_{n,j}$ and $\lambda_h(1-\sum_{j=\ell}^L T_{n,j})$, respectively.

Next, we calculate the conditional probability $\Pr[{\rm SIR}_{{\rm SVC},n,\ell}\geq \tau|d_0=x]$:
\begin{align}
\Pr[{\rm SIR}_{{\rm SVC},n,\ell}\geq \tau|d_0=x]= \mathbb E_{I_{n,\ell},\overline{I}_{n,\ell}}\left[\exp(-s(I_{n,\ell}+\overline{I}_{n,\ell}))\right]=\mathcal L_{I_{n,\ell}}(s) \mathcal L_{\overline{I}_{n,\ell}}(s),\label{eqn:conditional_SIR}
\end{align}
where $s=\tau d_0^{\alpha}$ and $\mathcal L_{I}(s)$ denotes the Laplace transform of random variable $I$.
To calculate $\Pr[{\rm SIR}_{{\rm SVC},n,\ell}\geq \tau|d_0=x]$, we first calculate $\mathcal L_{I_{n,\ell}}(s)$ and $\mathcal L_{\bar{I}_{n,\ell}}(s)$, respectively. $\mathcal L_{I_{n,\ell}}(s)$ can be calculated as follows:
\begin{align}\label{eqn:L_I_n}
&\mathcal{L}_{I_{n,\ell}}(s)={\mathbb E}\left[\exp\left(-s\sum_{i\in\Phi_{h,n,\ell}\backslash B_{{\rm SVC},n,\ell,0}}d_{i}^{-\alpha}\left|h_{i}\right|^{2} \right)\right]
={\mathbb E}\left[\prod_{i\in\Phi_{h,n,\ell}\backslash B_{{\rm SVC},n,\ell,0}}\exp\left(-s d_{i}^{-\alpha}\left|h_{i}\right|^{2} \right)\right]\notag\\
&\eqla\exp\left(-2\pi \Big(\sum_{j=\ell}^L T_{n,j}\Big)\lambda_{h}\int_{d_0}^{\infty}\left(1-\frac{1}{1+sr^{-\alpha}}\right)r{\rm d}r\right)\notag\\
&\eqlb\exp\left(-\frac{2\pi}{\alpha}\Big(\sum_{j=\ell}^L T_{n,j}\Big)\lambda_{h}s^{\frac{2}{\alpha}}B^{'}\left(\frac{2}{\alpha},1-\frac{2}{\alpha},\frac{1}{1+sd_0^{-\alpha}}\right)\right),
\end{align}
where (a) is obtained by using the probability generating functional of a PPP, (b) is obtained by first replacing $s^{-\frac{1}{\alpha}}r$ with $t$, and then replacing $\frac{1}{1+t^{-\alpha}}$ with $w$. Similar to $\mathcal L_{I_{n,\ell}}(s)$, we have:
\begin{align}
\mathcal L_{\overline{I}_{n,\ell}}(s)=\exp\left(-\frac{2\pi}{\alpha}\Big(1-\sum_{j=\ell}^L T_{n,j}\Big)\lambda_hs^{\frac{2}{\alpha}}B\left(\frac{2} {\alpha},1-\frac{2}{\alpha}\right)\right).\label{eqn:L_I_m}
\end{align}
Substituting \eqref{eqn:L_I_n} and \eqref{eqn:L_I_m} into \eqref{eqn:conditional_SIR}, we can obtain $\Pr[{\rm SIR}_{{\rm SVC},n,\ell}\geq \tau|d_0=x]$.

Finally, we calculate $\Pr[{\rm SIR}_{{\rm SVC},n,\ell}\geq \tau]$. Note that the p.d.f. of $d_0$ is $f_{d_0}(x)=2\pi\sum\limits_{j=\ell}^L T_{n,j}\lambda_h x\exp(-\pi\\ \times\sum_{j=\ell}^L T_{n,j}\lambda_hx^2)$. Thus, we have $\Pr[{\rm SIR}_{{\rm SVC},n,\ell}\geq \tau]=\int_{0}^{\infty}\Pr[{\rm SIR}_{{\rm SVC},n,\ell}\geq\tau|d_0=x]f_{d_0}(x){\rm d}x= \frac{\sum_{j=\ell}^L T_{n,j}}{D_2(\tau)+D_1(\tau)\sum_{j=\ell}^L T_{n,j}}$. Therefore, we complete the proof.

\section*{Appendix C: Proof of Lemma~\ref{Lem:svc_property_opt_T}}
First, we show Property (i) of Lemma~\ref{Lem:svc_property_opt_T}.
Consider the case of $\ell=1$. 
Suppose $\frac{b_{n,1}}{s_1}\leq \frac{b_{n,2}}{s_{2}}$ and $T_{{\rm SVC},n,1}^*>0$. Since $T_{{\rm SVC},n,1}^*>0$ and $T_{{\rm SVC},n,1}^*\lambda_{n,1}^*=0$, we have $\lambda^*_{n,1}=0$. Combining with $\eta_n^*\geq 0$, $\lambda_{n,2}^*\geq 0$ and  $\frac{b_{n,1}}{s_1}\leq \frac{b_{n,2}}{s_{2}}$, we have $T_{{\rm SVC},n,1}^*=\frac{1}{D_1(\tau_C)}\left(\sqrt{\frac{a_{n}b_{n,1}D_2(\tau_C)}{v^*s_1-\lambda_{n,1}^*+\eta_n^*}}-\sqrt{\frac{a_{n}b_{n,2}D_2(\tau_C)}{v^*s_2-\lambda_{n,2}^*+\lambda_{n,1}^*}}\right)\leq 0$, which contradicts the assumption. Thus, by contradiction, we can prove  $T_{{\rm SVC},n,1}^*=0$ if $\frac{b_{n,1}}{s_1}\leq \frac{b_{n,2}}{s_{2}}$.
Consider the case of $\ell\in\{2,\cdots,L-1\}$. Suppose $\frac{b_{n,\ell}}{s_\ell}\leq \frac{b_{n,\ell+1}}{s_{\ell+1}}$ and $T_{{\rm SVC},n,\ell}^*>0$. Since $T_{{\rm SVC},n,\ell}^*>0$ and $T_{{\rm SVC},n,\ell}^*\lambda_{n,\ell}^*=0$, we have $\lambda^*_{n,\ell}=0$. Combining with $\lambda_{n,\ell-1}^*\geq 0$, $\lambda_{n,\ell+1}^*\geq 0$ and $\frac{b_{n,\ell}}{s_\ell}\leq \frac{b_{n,\ell+1}}{s_{\ell+1}}$, we have $T_{{\rm SVC},n,\ell}^*=\frac{1}{D_1(\tau_C)}\left(\sqrt{\frac{a_{n}b_{n,\ell}D_2(\tau_C)}{v^*s_\ell-\lambda_{n,\ell}^*+\lambda_{n,{\ell-1}}^*}}-\sqrt{\frac{a_{n}b_{n,{\ell+1}}D_2(\tau_C)}{v^*s_{\ell+1}-\lambda_{n,{\ell+1}}^*+\lambda_{n,{\ell}}^*}}\right)\leq 0$, which contradicts the assumption. Thus, by contradiction, we can prove  $T_{{\rm SVC},n,\ell}^*=0$ if $\frac{b_{n,\ell}}{s_\ell}\leq \frac{b_{n,\ell+1}}{s_{\ell+1}}$.
Therefore, we prove property (i) of Lemma~\ref{Lem:svc_property_opt_T}.

Next, we show Property (ii) of Lemma~\ref{Lem:svc_property_opt_T}. Since $T_{{\rm SVC},n,j}^*>0$  and $T_{{\rm SVC},n,j}^*\lambda_{n,j}^*=0$ for all $j\in\mathcal L$, we have $\lambda_{n,j}^*=0$ for all $j\in\mathcal L$. Thus, $T_{{\rm SVC},n,\ell}^*$ can be rewritten as
\begin{align}
&T_{{\rm SVC},n,\ell}^*=\begin{cases}
\frac{1}{D_1(\tau_C)}\left(\sqrt{\frac{a_{n}b_{n,1}D_2(\tau_C)}{v^*s_1+\eta_n^*}}-\sqrt{\frac{a_{n}b_{n,2}D_2(\tau_C)}{v^*s_2}}\right), &\quad \ell=1\\
\frac{1}{D_1(\tau_C)}\left(\sqrt{\frac{a_{n}b_{n,\ell}D_2(\tau_C)}{v^*s_\ell}}-\sqrt{\frac{a_{n}b_{n,{\ell+1}}D_2(\tau_C)}{v^*s_{\ell+1}}}\right), &\quad \ell\in\{2,\cdots,L-1\}\\
\frac{1}{D_1(\tau_C)}\sqrt{\frac{a_{n}b_{n,L}D_2(\tau_C)}{v^*s_L}}-\frac{D_2(\tau_C)}{D_1(\tau_C)},  &\quad \ell=L
\end{cases}. \notag
\end{align}
Consider the case of $\ell=1$. If $\sqrt{\frac{b_{n,2}}{s_{2}}}-\sqrt{\frac{b_{n,3}}{s_{3}}}\geq \sqrt{\frac{b_{n,1}}{s_{1}}}-\sqrt{\frac{b_{n,2}}{s_{2}}}$, we have
\begin{align}
T_{{\rm SVC},n,2}^*&=\sqrt{\frac{a_{n}D_2(\tau_C)}{v^*(D_1(\tau_C))^2}}\left(\sqrt{\frac{b_{n,2}}{s_2}}-\sqrt{\frac{b_{n,{3}}}{s_{3}}}\right)\geq \sqrt{\frac{a_{n}D_2(\tau_C)}{v^*(D_1(\tau_C))^2}}\left(\sqrt{\frac{b_{n,1}}{s_1}}-\sqrt{\frac{b_{n,{2}}}{s_{2}}}\right)\notag\\
&\stackrel{(a)}{\geq} \sqrt{\frac{a_{n}D_2(\tau_C)}{v^*(D_1(\tau_C))^2}}\left(\sqrt{\frac{b_{n,1}}{s_1+\frac{\eta_n^*}{v^*}}}-\sqrt{\frac{b_{n,{2}}}{s_{2}}}\right) =T_{{\rm SVC},n,1}^*,\notag
\end{align}
where (a) is due to $\eta_n^*\geq 0$. Consider the case of $\ell\in\{2,\cdots,L-2\}$. If $\sqrt{\frac{b_{n,\ell+1}}{s_{\ell+1}}}-\sqrt{\frac{b_{n,\ell+2}}{s_{\ell+2}}}\geq \sqrt{\frac{b_{n,\ell}}{s_{\ell}}}-\sqrt{\frac{b_{n,\ell+1}}{s_{\ell+1}}}$, we have
\begin{align}
T_{{\rm SVC},n,\ell+1}^*=\sqrt{\frac{a_{n}D_2(\tau_C)}{v^*(D_1(\tau_C))^2}}\left(\sqrt{\frac{b_{n,\ell+1}}{s_{\ell+1}}}-\sqrt{\frac{b_{n,{\ell+2}}}{s_{\ell+2}}}\right)&\geq \sqrt{\frac{a_{n}D_2(\tau_C)}{v^*(D_1(\tau_C))^2}}\left(\sqrt{\frac{b_{n,\ell}}{s_{\ell}}}-\sqrt{\frac{b_{n,{\ell+1}}}{s_{\ell+1}}}\right)\notag\\
&=T_{{\rm SVC},n,1}^*.\notag
\end{align}
Therefore, we prove property (ii) of Lemma~\ref{Lem:svc_property_opt_T}.

Finally, we show Property (iii) of Lemma~\ref{Lem:svc_property_opt_T}. For any $n\in\mathcal N$ with $T_{{\rm SVC},	n,j}^*>0$, $j\in\mathcal L$, we have
\begin{align}
\sum_{i=\ell}^L T_{{\rm SVC},n,i}^*= \begin{cases}
\frac{1}{D_1(\tau_C)}\sqrt{\frac{a_{n}b_{n,1}D_2(\tau_C)}{v^*s_1+\eta_n^*}}-\frac{D_2(\tau_C)}{D_1(\tau_C)}, &\quad \ell=1\\
\frac{1}{D_1(\tau_C)}\sqrt{\frac{a_{n}b_{n,\ell}D_2(\tau_C)}{v^*s_\ell}}-\frac{D_2(\tau_C)}{D_1(\tau_C)}, &\quad \ell \in\{2,\cdots,L\}
\end{cases}.\notag
\end{align}
For any $n_1,n_2\in\mathcal N$ with $T_{{\rm SVC},n_1,j}^*,T_{{\rm SVC},n_2,j}^*>0$, $j\in\mathcal L$, consider the following two cases:
(a) For $\ell=1$ with $a_{n_1}b_{n_1,1}\geq a_{n_2}b_{n_2,1}$, if $\eta_{n_1}^*>0$, we have $\sum_{i=1}^LT_{{\rm SVC},n_1,i}^*=1\geq \sum_{i=1}^LT_{{\rm SVC},n_2,i}^*$; if $\eta_{n_1}^*=0$, we have $\sum_{i=1}^LT_{{\rm SVC},n_1,i}^*=\frac{1}{D_1(\tau_C)}\sqrt{\frac{a_{n_1}b_{n_1,1}D_2(\tau_C)}{v^*s_1}}-\frac{D_2(\tau_C)}{D_1(\tau_C)}\geq \frac{1}{D_1(\tau_C)}\sqrt{\frac{a_{n_2}b_{n_2,1}D_2(\tau_C)}{v^*s_1+\eta_{n_2}^*}}-\frac{D_2(\tau_C)}{D_1(\tau_C)}= \sum_{i=1}^LT_{{\rm SVC},n_2,i}^*$.
(b) For any $\ell\in\{2,\cdots,L\}$ with $a_{n_1}b_{n_1,\ell}\geq a_{n_2}b_{n_2,\ell}$, we have $\sum_{i=\ell}^LT_{{\rm SVC},n_1,i}^*=\frac{1}{D_1(\tau_C)}\sqrt{\frac{a_{n_1}b_{n_1,\ell}D_2(\tau_C)}{v^*s_\ell}}-\frac{D_2(\tau_C)}{D_1(\tau_C)}\geq \frac{1}{D_1(\tau_C)}\sqrt{\frac{a_{n_2}b_{n_2,\ell}D_2(\tau_C)}{v^*s_\ell}}-\frac{D_2(\tau_C)}{D_1(\tau_C)}=\sum_{i=\ell}^LT_{{\rm SVC},n_2,i}^*$.
Therefore, we prove property (iii) of Lemma~\ref{Lem:svc_property_opt_T}.



\end{document}